\documentclass[prl,twocolumn,showpacs,preprintnumbers,amsmath,amssymb,superscriptaddress,letterpaper]{revtex4-1}

\usepackage{amssymb}
\usepackage{amsmath}
\usepackage{balance} 
\usepackage{color} 

\usepackage{graphicx} %eps figures can be used instead
\usepackage{multirow}
\usepackage{color}
\usepackage{booktabs}

\begin{document}

\title{Ceria co-doping: Synergistic or average effect?}

\author{Mario Burbano}
\affiliation{School of Chemistry and CRANN, Trinity College Dublin, Dublin 2, Ireland}
\author{Sian Nadin}
\affiliation{School of Chemistry and CRANN, Trinity College Dublin, Dublin 2, Ireland}
\author{Dario Marrocchelli}
\affiliation{Department of Nuclear Science and Engineering, Massachusetts Institute of Technology, Cambridge, MA, United States of America}
\author{Mathieu Salanne}
\affiliation{Sorbonne Universit\'{e}s, UPMC Univ Paris 06, CNRS, UMR 8234, PHENIX, F-75005 Paris, France}
\author{Graeme W. Watson}
\affiliation{School of Chemistry and CRANN, Trinity College Dublin, Dublin 2, Ireland}

%Please do not change this text.

\begin{abstract}
Ceria (CeO$_{2}$) co-doping has been suggested as a means to achieve ionic conductivities that are \emph{significantly} higher than those in singly-doped systems. Rekindled interest in this topic over the last decade has given rise to claims of much improved performance. The present study makes use of computer simulations to investigate the bulk ionic conductivity of Rare Earth (RE) doped ceria, where RE = Sc, Gd, Sm, Nd and La. The results from the singly doped systems are compared to those from ceria co-doped with Nd/Sm and Sc/La. The pattern that emerges from the conductivity data is consistent with the dominance of local lattice strains from individual defects, rather than the synergistic co-doping effect reported recently and, as a result no enhancement in the conductivity of co-doped samples is observed.
\end{abstract}

\maketitle

%additional addresses can be cited as above using the lower-case letters, c, d, e... If all authors are from the same address, no letter is required

%\footnotetext{\ddag~Additional footnotes to the title and authors can be included \emph{e.g.}\ `Present address:' or `These authors contributed equally to this work' as above using the symbols: \ddag, \textsection, and \P. Please place the appropriate symbol next to the author's name and include a \texttt{\textbackslash footnotetext} entry in the the correct place in the list.}

\section{Introduction}
\label{intro}
Distributed Generation of electricity has been touted by national and international agencies, such as, the Organisation for Economic Co-operation and Development (OECD) \cite{KargerAndHennings_RSER2009,DE_IEA2002,CHP_IEA2008}, as a possible means to incorporate nascent technologies into the energy market. Some of these technologies rely on renewable sources, e.g. wind and solar, but they are still severely limited by high production costs. For this reason,  more established technologies  continue to attract significant attention. One such technology is called Combined Heat and Power (CHP) where the excess heat generated in electricity production is recycled to improve the overall efficiency of the process. CHP systems have traditionally made use of steam turbines, gas turbines, reciprocating engines and, more recently, Solid Oxide Fuel Cells (SOFCs) \cite{Ormerod_CSR2003}. The latter are particularly interesting given their high efficiency (up to 85\%), low pollutant emissions and fuel flexibility (they can utilize hydrogen, natural gas, landfill gas, gasified coal, etc \cite{GuoEtAl_EES2013}); thus, SOFCs have the potential to play a key role in the energy conversion landscape for the medium and long term future \cite{WachsmanEtAl_EES2012,WachsmanAndLee_S2011,QinEtAl_EES2011,Ruiz-MoralesEtAl_ESS2010}. Commercial applications of SOFCs could range from domestic units to small power stations. Nonetheless, widespread use of SOFCs has been held back by their high operating temperature, which requires the use of expensive materials and affects the long-term performance of these devices. Substantial research efforts have been devoted to lowering their operating regime to between 500 to 750 C$^\circ$, known as the intermediate temperature (IT) range \cite{BrettEtAl_CSR2008}. There are two processes which limit significantly the performance of IT-SOFCs, namely the Oxygen Reduction Reaction (ORR) at the cathode and the ionic conductivity of the electrolyte. This article focuses on the latter effect as a means to improve the performance of SOFC electrolytes, in particular those based on doped ceria (CeO$_{2}$).
\newline

Fluorite-structured materials, such as CeO$_2$, ZrO$_2$ and $\delta$-Bi$_2$O$_3$ are among the best oxide ion conductors, which makes them ideal candidates for use as electrolytes in IT-SOFCs. The conduction mechanism in these ceramics is known to occur by means of vacancy migration in the anion sublattice \cite{TullerAndNowick_JES1975,TullerAndNowick_JPCS1977}. Oxygen vacancies are introduced by doping these materials with lower valent cations\footnote{$\delta$-Bi$_2$O$_3$ actually behaves differently, since this material already has 25\% intrinsic vacancies. In this case doping with iso-valent cations is used to stabilise the fluorite structure at low temperatures \cite{AbrahamsEtAL_CM2010}.}. For example, CeO$_2$ is usually doped with Rare Earth cations (RE = lanthanides + Sc and Y) which leads to the formation of one vacancy ($\mathrm{V^{\cdot\cdot}_{O}}$) for each pair of cations, as illustrated by Equation \ref{eq:vac_form}, in Kr\"oger-Vink notation \cite{SunEtAl_EES2012}:

\begin{equation}
\mathrm{RE_{2}O_{3} + 2Ce_{Ce}^{x} + O_{O}^{x} \rightarrow 2RE^{'}_{Ce} + V^{\cdot\cdot}_{O} + 2CeO_{2}}
\label{eq:vac_form}
\end{equation} 

The formation of oxygen vacancies has a beneficial effect on the conductivity, which shows, at first, a marked increase as more vacancies are added. This increase, however, does not display a monotonic behaviour, but rather, a sharp drop in conductivity is observed beyond a critical concentration. In the case of ceria, this value varies between 2.5 and 5 \% vacancy concentration, depending on the dopant cation species, the temperature and other factors. There is a general consensus that this drop in conductivity with increased number of vacancies is caused by defect--defect interactions \cite{BogicevicEtAl_PRB2001,BogicevicAndWolverton_PRB2003,NavrotskyEtAl_FD2007,PietrucciEtAl_PRB2008,Navrotsky_JMC2010,NorbergEtAl_CM2011,MarrocchelliEtAl_CM2011,BurbanoEtAl_CM2012,ChenEtAl_CM2012,MarrocchelliEtAl_JMCA2013}. The interactions between these defects can be broadly classified into cation-vacancy, vacancy-vacancy and cation-cation \cite{BogicevicEtAl_PRB2001,BogicevicAndWolverton_PRB2003,NavrotskyEtAl_FD2007,PietrucciEtAl_PRB2008,Navrotsky_JMC2010,NorbergEtAl_CM2011,MarrocchelliEtAl_CM2011,BurbanoEtAl_CM2012}.  In practice, these interactions can prove deleterious to the ionic conductivity because defect association leads to fewer mobile vacancies being available \cite{ChenEtAl_CM2012}, thus the need for high operating temperatures. This process is generally summarised by the following formula for the conductivity, $\sigma$:
\begin{equation}
\sigma T= \sigma_0 \;\exp\left(-\frac{E_a}{k_BT}\right),
\label{eq:sigma}
\end{equation} 
where $T$ is the temperature, $\sigma_0$ the composition-dependent pre-exponential factor, $E_a$ the activation energy and $k_B$ the Boltzmann constant. The activation energy is given by the sum of a migration enthalpy, $\Delta H_m$, and a defect association enthalpy $\Delta H_{ass}$. The higher the association enthalpy, the lower the conductivity.
\newline

Traditionally, ionic conductivity optimization has been approached from a compositional perspective. This has meant that, in order to improve the conductivity of ceria-based electrolytes, researchers have mostly focused on parameters such as the ionic radius of the dopant cation and its concentration, ${x}$ in Ce$_{1-x}$RE$_{x}$O$_{2-x/2}$. The chief aim has been to minimize defect interactions, especially those between dopants and vacancies \cite{BogicevicAndWolverton_PRB2003,ButlerEtAl_SSI1983,Kilner_SSI1983,BalazsAndGlass_SSI1995,HayashiEtAL_SSI2000,AnderssonEtAl_PNAS2006}. These studies have variously identified a number of RE elements, such as Gd$^{3+}$, Y$^{3+}$, Sm$^{3+}$ and Pm$^{3+}$ as the best candidate dopants, given that their radius mismatch with the host cation (Ce$^{4+}$) balances the competing electrostatic and elastic components of the defect interactions which control their association \cite{WangEtAl_AC2011}. Nevertheless, recent studies have shown that in the limit where cation-vacancy interactions are reduced to a minimum, it is vacancy-vacancy association which ultimately determines the ionic conductivity drop as a function of dopant concentration in fluorite-structured materials, such as, Yttria Doped Ceria (YDC) \cite{BurbanoEtAl_CM2012}, Yttria Stabilized Zirconia (YSZ) and Scandia Stabilized Zirconia (ScSZ) \cite{MarrocchelliEtAl_CM2011}. This means that different optimization strategies must be sought if IT-SOFCs are to realize their potential in commercial applications \cite{WachsmanEtAl_EES2012,WachsmanAndLee_S2011,Rupp_SSI2012,BurbanoEtAl_JECR2013}.  \newline

An interesting route for improving the ionic conductivity of fluorite-structured electrolytes is co-doping, i.e. doping these materials with more than one cation species. This approach was employed on ZrO$_{2}$ by Politova and Irvine \cite{PolitovaAndIrvine_SSI2004} with two different cation species, each playing a different role. In this case, the material was doped with two different cation species, where each of which played a {\em different} role. Sc was added to improve the ionic conductivity, since its radius is very close to that of Zr, thus minimizing cation-vacancy interactions; while Y, on the other hand, was introduced because its larger ionic radius fully stabilizes the fluorite structure and removes a phase transition to a lower-symmetry phase observed in pure Sc-doped ZrO$_{2}$. It is important to note that Y addition to Sc-doped ZrO$_{2}$ \emph{lowers} the conductivity of this material. Therefore, a {\em compromise} exists between stability and ionic conductivity in this co-doped zirconia system. In the case of Politova and Irvine, the authors found that very \textit{small} concentrations of Y are necessary to stabilize the cubic fluorite structure and that this has a small effect on the conductivity.
\newline

It has been suggested that co-doping can also be used to improve the ionic conductivity of ceria based electrolytes~\cite{vanHerleEtAl_SECS1999,SinghEtAl_Ionics2013,SinghEtAl_Ionics2012,OmarEtAl_JACS2009,OmarEtAl_APL2007,ShaEtAl_JAC2006,ShaEtAl_JAC2007,DikmenEtAl_JPS2010,GuanEtAl_MRB2008,AyawannaEtAl_SSI2009,DholabhaiEtAl_JMC2011,RalphEtAl_BBG1997,KasseAndNino_JAC2013}. However, contrary to the stabilizing role it plays in zirconia, co-doping in ceria has been used in order to either reproduce the ionic radius of an ideal dopant, or the lattice constant of ceria doped with said dopant. Co-doping with two or more different cations aims to obtain an {\em average} or ``effective'' cation radius that is very close to that of Ce$^{4+}$, hence the average strain introduced by the dopant cations is minimized. This is substantiated by different interpretations of how a dopant with a critical radius ($r_{\mathrm{C}}$) is likely to affect defect--defect interactions. For example, in 1989 Kim suggested that the ideal dopant would not change the volume of the host lattice upon its introduction, thus minimizing the elastic strain, and identified $r_\mathrm{C}$ = 1.038\,\AA\ \cite{Kim_JACS1989}. More recently, researchers have had access to a more detailed view of the interplay between strain and electrostatics with the use of \textit{ab initio} methods, which in 2006 lead Andersson \textit{et al.} \cite{AnderssonEtAl_PNAS2006} to ascertain that $r_{\mathrm{C}}$ should be that which maximizes oxygen vacancy disorder. Their simulations showed that when ceria is doped with relatively small cations, vacancies prefer to sit in a nearest neighbour (NN) position with respect to the dopant cation, whereas, for larger cations, vacancies prefer to sit in a next nearest neighbour (NNN) position. The crossover between these two tendencies was observed at Pm$^{3+}$, for which the NN and NNN positions have the same energy, which was rationalized in terms of a perfect balance between the elastic (related to the dopant's radius) and Coulombic interactions between Pm$^{3+}$ and a vacancy. This finding implies that Pm$^{3+}$ is the ideal dopant for ceria because it increases the configurational entropy and should display the highest ionic conductivity. Unfortunately, Pm$^{3+}$ is radioactive, so the authors suggested, instead, to try a mixture of Sm/Nd which have slightly smaller/larger ionic radii than Pm$^{3+}$.\footnote{We note here that the radii of these three elements (1.079, 1.093, 1.109 \AA\ for Sm, Pm and Nd, respectively \cite{Shannon_ACA1976}) are all very similar and almost within the associated experimental error.} Based on these ideas, multiple research groups have carried out experiments in order to test several co-doping schemes for ceria, e.g. Y/Sm co-doping~\cite{ShaEtAl_JAC2006}, La/Y co-doping~\cite{ShaEtAl_JAC2007}, Lu/Nd co-doping~\cite{OmarEtAl_SSI2006} and Sm/Nd co-doping~\cite{OmarEtAl_APL2007}, of which the latter two were specifically aimed at reproducing the $r_\mathrm{C}$ values predicted by Kim and by Andersson, respectively. In general, co-doping studies have pointed to increases in the ionic conductivity with respect to singly doped systems, which has lead to the conclusion that there exists a ``co-doping'' effect in ceria. 
\newline

Modern simulation techniques have become a mainstay within the materials science community, not only because they afford researchers information which is complementary to their experiments, but also because they can serve as predictive tools \cite{ChroneosEtAl_EES2011}. Hence, the implementation of reliable computer simulations can be used to clarify the role of particular effects present in physical experiments in a targeted and controlled manner. To this end, the interaction potentials reported here are shown to perform with the accuracy of state-of-the-art {\it first-principles} calculations, i.e. hybrid Density Functional Therory (DFT), but at the computational cost of classical (polarizable) molecular dynamics. This approach allows us to study systems with realistic doped/co-doped defect concentrations within the temperatures of interest for SOFC applications (600-1000 C$^\circ$), and to accumulate sufficiently long trajectories to calculate the conductivity. This is in contrast to most of the previous computational work on doped ceria which has typically used static DFT calculations or emprirical potentials fitted to equilibrium properties \cite{ButlerEtAl_SSI1983,BalducciEtAl_CM2000}. The use of computer simulations allows us to focus on the \emph{bulk} behaviour of this material, excluding factors like grain size and boundaries, sintering conditions, impurity levels, etc, which are known to also (negatively) affect the conductivity of these materials \cite{Goodenough_ARMR2003,GobelEtAl_PCCP2010}. We show that co-doping does not significantly improve the conductivity of these materials, but rather, we find that the conductivity of the co-doped systems lies within the range spanned by the singly doped systems, i.e. it is an average of the two. The reason for this is that introducing two cation species with radii which are bigger or smaller than that of a given $r_{\mathrm{C}}$ affects the local structure of ceria and results in deep traps for the vacancies.
\newline

\section{Methods}
\label{Methods}
\subsection{Interionic Potential} 
The highly correlated nature of the $f$-electrons found in lanthanide elements makes necessary the use of high levels of theory in order to correctly describe their electronic structure. Such demands have been found to be satisfied by the inclusion of a fraction of non-local Hartree-Fock exchange within the framework of Density Funcional Theory (DFT), which gives rise to hybrid functionals (h-DFT) \cite{DaSilvaEtAl_PRB2007,GillenEtAl_PRB2013,Ganduglia-PirovanoEtAl_SSR2007}. Alternative DFT functionals are also available, namely, those which include a Hubbard parameter $U$ (DFT$+U$). They represent a viable alternative to h-DFT and their ability to describe the properties of ceria has been widely documented \cite{KeatingEtAl_JPCM2009,KeatingEtAl_JPC2012,KeatingEtAl_JMCC2013,NolanEtAL_SSI2006,AnderssonEtAl_PhysRevB2007}. Nonetheless, h-DFT provides a better agreement with experimental lattice constants and does not require fitting a $+U$ value for the $f$-electron systems. In either case, however, DFT calculations are prohibitively expensive from a computational point of view for this type of study regardless of the functional; this is because of the long Molecular Dynamics (MD) simulation times and large systems required to study the ionic conductivity of doped ceria. For this reason interionic potentials (IP) implemented in an in-house MD code (PIMAIM) \cite{MaddenAndWilson_CSR1996} and derived from static h-DFT calculations were used in this work, as they accurately reproduce the structural \textit{ab initio} data at a fraction of the computational cost. This approach has been successfully used for a series of related oxides \cite{MarrocchelliEtAl_AFM2012,Marrocchelli_JPCM2010,MarrocchelliEtAl_MP2009,MarrocchelliEtAl_CM2011,MarrocchelliEtAl_JPCM2009,NorbergEtAl_CM2011,WilsonEtAl_JPCM2004}, including Y-doped ceria \cite{BurbanoEtAl_JPCM2011,BurbanoEtAl_CM2012}, as well as a variety of ionic systems \cite{HeatonEtAl_JPCB2006,SalanneEtAl_JFC2009,SalanneEtAl_JPCC2012}. The RE dopant cations studied in this article included La, Nd, Sm and Gd, as well as, Sc. A crucial feature of this potential set is that they were fitted with a \emph{common} O -- O term, which made it possible to perform simulations with several dopant cations within the same cell, i.e. to co-dope ceria. Details on the interionic potential used (DIPPIM - DIPole Polarizable Ionic Model), its parameterization and the parameters used are found in Appendix \ref{DIPPIM} and Appendix \ref{parameterization}, respectively.

\subsection{Simulation Details}
\label{sect:simus}
All MD simulations on the singly doped Ce$_{1-x}$RE$_{x}$O$_{2-x/2}$ and co-doped systems were performed using 6 $\times$ 6 $\times$ 6 supercells ($\sim$ 2592 atoms, depending on the dopant concentration). Two different co-doped systems formed part of this study. Firstly, Ce$_{1-x}$Nd$_{0.5x}$Sm$_{0.5x}$O$_{2-x/2}$, which has been previously studied experimentally \cite{OmarEtAl_APL2007} because the average radius of both dopants matches that of Pm. Similarly, Ce$_{1-x}$Sc$_{0.22x}$La$_{0.78x}$O$_{2-x/2}$ was included given that this ratio of La and Sc also reproduces the ionic radius of Pm. Three different supercells were set up for each dopant concentration, $x$ = 0.05, 0.10, 0.15, 0.20, 0.25 and the values reported here, such as, ionic conductivities, lattice constants and activation energies were obtained from the averages of these three configurations. Each calculation was set up by randomly distributing the dopants over the cation sublattice and the oxygen vacancies over the anion sublattice. The supercells were initially equilibrated at a temperature of 1673\,K for 40 ps; the temperature was then scaled down to room temperature at a rate of 2\,K ps$^{-1}$. The diffusion coefficients were calculated for temperatures between 873\,K and 1473\,K from simulations that were up to 3\,ns long in the case of the lowest temperature. The 300\,K lattice constants were obtained from 10 ps long runs. All simulations were performed at constant temperature and pressure (NPT ensemble), as described by Martyna et al. \cite{MartynaEtAl_JCP1994} using a time step of 1 fs. The Coulombic and dispersion interactions were summed using Ewald summations \cite{Ewald_AP1921}, while the short-range part of the potential was truncated at 12.96 \AA.

\section{Results and discussion}
\label{Results}

The reliability of the models used in the computer simulations presented throughout this work is assessed in Section \ref{Assesment} by means of comparison against experimental and computational results for singly doped ceria. We also note that this approach has already been successfully used to model yttria-doped ceria, as reported in ref. \cite{BurbanoEtAl_JPCM2011}. Section \ref{Co-doped} builds upon these results to determine whether co-doping is likely to improve the ionic conductivity in these solid electrolytes. The remainder of this article abbreviates the various RE-Doped Ceria systems (Ce$_{1-x}$RE$_{x}$O$_{2-x/2}$) under study to ScDC, GdDC, SmDC, NdDC and LaDC. In a similar fashion, the co-doped systems are referred to as Sc:LaDC and Nd:SmDC. 

\subsection{Potential Assessment}
\label{Assesment}
\subsubsection{Lattice constants of singly doped ceria}
\begin{figure*}[htbp]
\begin{center}
\includegraphics[scale=0.4]{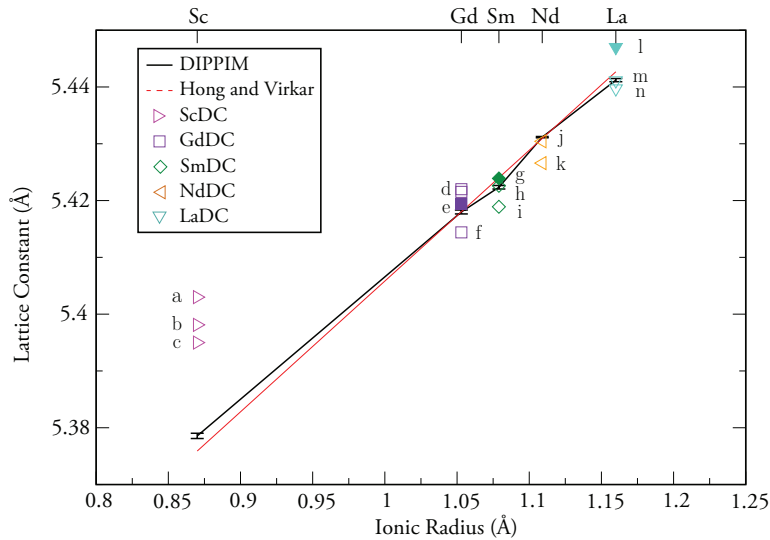}
\end{center}
\caption{Lattice constants for Ce$_{0.90}$RE$_{0.10}$O$_{1.95}$. DIPPIM values at 300\,K (this work) are shown as a solid black line. The errors correspond to the standard deviation obtained from three simulations. Hong and Virkar \cite{HongAndVirkar_JACS1995} values are represented by a dashed red line. Literature data from a) Grover \textit{et al.}~\cite{GroverEtAl_JSSC2008}, b) Lee \textit{et al.}~\cite{LeeEtAl_JSEE2012}, c) Gerhardt-Anderson and Nowick~\cite{Gerhardt-AndersonAndNowick_SSI1981}, d) Huang \textit{et al.}~\cite{HuangEtAl_JACS1998} and Zhang \textit{et al}~\cite{ZhangEtAl_SSI2002} e) Huang \textit{et al.}~\cite{HuangEtAl_JECS2011} f) Omar \textit{et al.}~\cite{OmarEtAl_JACS2009}, g) Huang \textit{et al.}~\cite{HuangEtAl_JECS2011}, h) Buyukkilic \textit{et al.}~\cite{BuyukkilicEtAl_SSI2012}, i) Omar \textit{et al.}~\cite{OmarEtAl_JACS2009}, j) Buyukkilic \textit{et al.}~\cite{BuyukkilicEtAl_SSI2012}, k) Omar \textit{et al.}~\cite{OmarEtAl_JACS2009}, l) Huang \textit{et al.}~\cite{HuangEtAl_JECS2011} m) Hisashige \textit{et al.}~\cite{HisashigeEtAl_JAC2006} and n) Dikmen \textit{et al.}~\cite{DikmenEtAl_SSI1999}}
\label{a0-300K}
\end{figure*}

\
\label{sect:a0-single}
The DIPPIM simulated lattice constants for Ce$_{0.90}$RE$_{0.10}$O$_{1.95}$ at 300\,K are shown in Figure \ref{a0-300K} as a solid black line. Their associated errors are represented by the standard deviation of the values from the three simulations carried out for each system. The dashed red line in Figure \ref{a0-300K} corresponds to the values for the same systems predicted by Hong and Virkar \cite{HongAndVirkar_JACS1995}, who  derived an empirical expression for the relationship between the ionic radius of the dopant and the lattice constant of RE-doped ceria. Figure \ref{a0-300K} also presents experimental (open symbols) and computational (closed symbols) lattice constant values for the same compositions (10\% cation doped) of ScDC, GdDC, SmDC, NdDC and LaDC.  

As shown in a previous study by the authors \cite{BurbanoEtAl_CM2012}, DIPPIM simulations are expected to perform as well as the DFT functional from which they were parameterized. In this case, the use of hybrid DFT functionals means that errors in the calculated value of lattice constants with respect to experiment for doped ceria should be in the order of 0.20\% \cite{DaSilvaEtAl_PRB2007}. This is borne out by the results presented in Figure \ref{a0-300K}, which show an excellent agreement with the range of experimental data available in the literature and also with the values calculated using Hong and Virkar's equation. In the case of ScDC, the simulations provide a better estimation of the lattice constant than that obtained from Hong and Virkar with respect to experiment, however both models underestimate the value of $a_{0}$. For DIPPIM simulations of ScDC, the largest error is 0.45\% compared to the value from Grover \textit{et al.} \cite{GroverEtAl_JSSC2008}. Despite this being a relatively small error, it is likely that this discrepancy arises from sources other than the DFT functional employed in this work. In fact, of the three experimental sources cited, Grover \textit{et al.} (highest value for a$_{0}$), as well as, Gerhardt-Anderson and Nowick \cite{Gerhardt-AndersonAndNowick_SSI1981} (lowest value for a$_{0}$) predict low solubilities for scandia (Sc$_{2}$O$_{3}$) and thus, formation of C-type phases in Ce$_{0.90}$Sc$_{0.10}$O$_{1.95}$ which are characteristic of sesquioxides that crystallize in the cubic bixbyite structure, such as scandia. This phase separation in Ce$_{0.90}$Sc$_{0.10}$O$_{1.95}$ was recently demonstrated by a series of elegant simulations that coupled DFT$+U$ and Monte Carlo simulations \cite{GrieshammerEtAl_PCCP2014}. 

\subsubsection{Ionic conductivity of singly doped ceria}
\begin{figure*}[htbp]
\begin{center}
\includegraphics[scale=0.4]{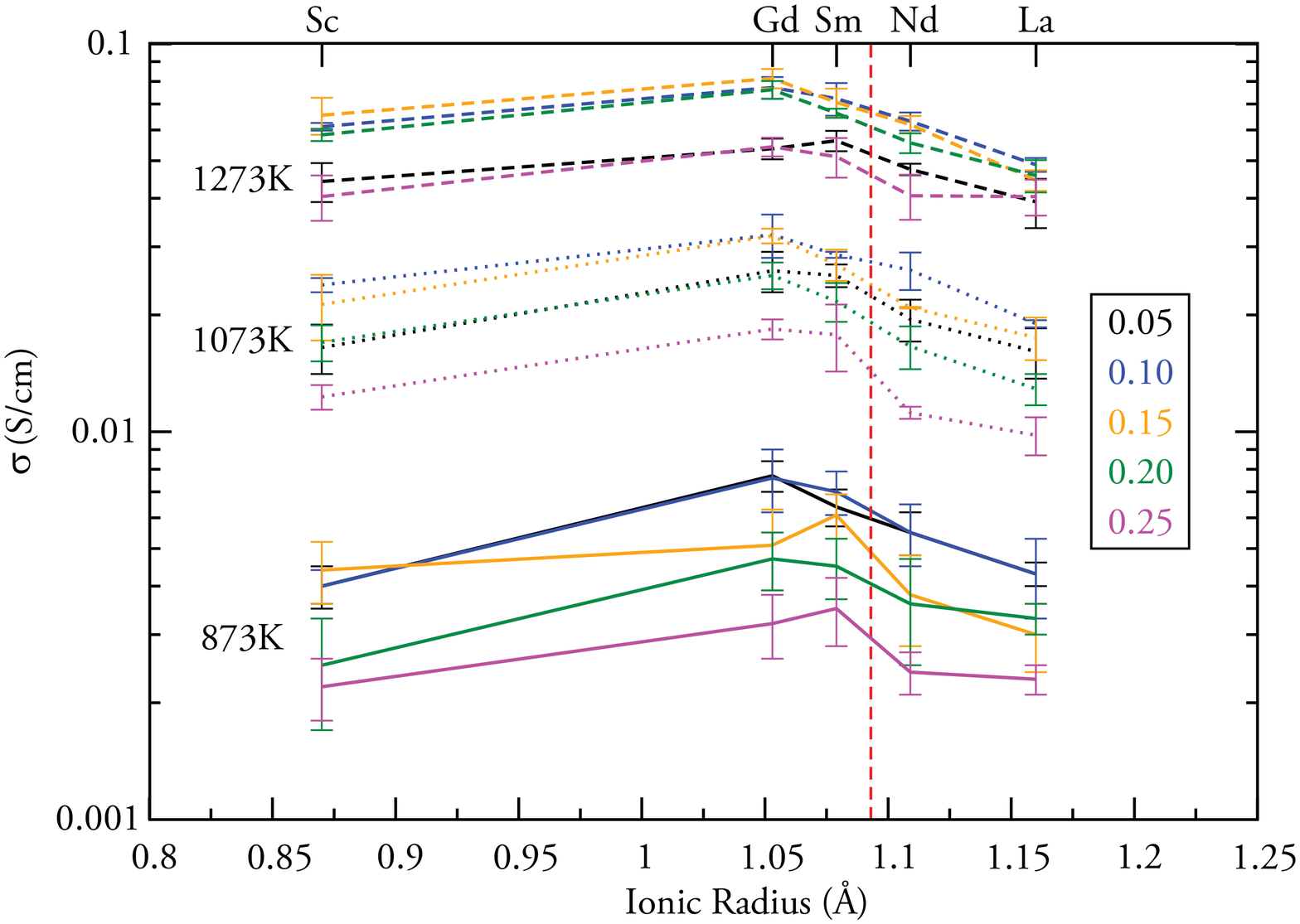}
\end{center}
\caption{Ionic conductivities for singly doped ceria (Ce$_{1-x}$RE$_{x}$O$_{2-x/2}$) at 1273\,K (blue diamonds), 1073\,K (black dots) and 873\,K (red triangles). The vertical dashed red line represents the critical ionic radius ($r_{\mathrm{C}}$) introduced by Andersson \textit{et al.} \cite{AnderssonEtAl_PNAS2006}.}
\label{SigAllTemps}
\end{figure*}

\ The ionic conductivities calculated for all the singly doped ceria systems under study are presented in Figure \ref{SigAllTemps}. The conductivities from simulations at 1273\,K are indicated by the dashed lines, those at 1073\,K by dotted lines and solid lines for those at 873\,K. These lines are to be interpreted only as a guide to the eye, as they connect the individual values from the calculations, each of which has an associated error indicated by the standard deviation from three measurements. The colours differentiate the concentration of the dopant cations as a percentage, thus 5\% is shown in black, 10\% in blue, 15\% in orange, 20\% in green and 25\% in magenta. In agreement with the literature for singly doped ceria, the best dopants were found to be Gd and Sm \cite{Steele_SSI2000,BuyukkilicEtAl_SSI2012,ButlerEtAl_SSI1983}. In particular, GdDC was found to have the highest conductivity at all temperatures. The results also show that the concentration which gives the highest conductivity varies from one doped system to another for a particular temperature. However 10\% GdDC is consistently among the systems that display the highest conductivity. In addition, the errors associated with each measurement become larger at lower temperatures because of the slower diffusion. As was mentioned above, Sc is soluble in ceria only in small amounts, hence the results presented here for a fluorite structure with randomly distributed cations correspond to an idealized description of ScDC. The dashed red line (vertical) shown in Figure \ref{SigAllTemps} represents Andersson's critical ionic radius \cite{AnderssonEtAl_PNAS2006} ($r_{\mathrm{c}}$) for the ideal dopant cation. \newline

\begin{figure*}[htbp]
\begin{center}
\includegraphics[scale=0.4]{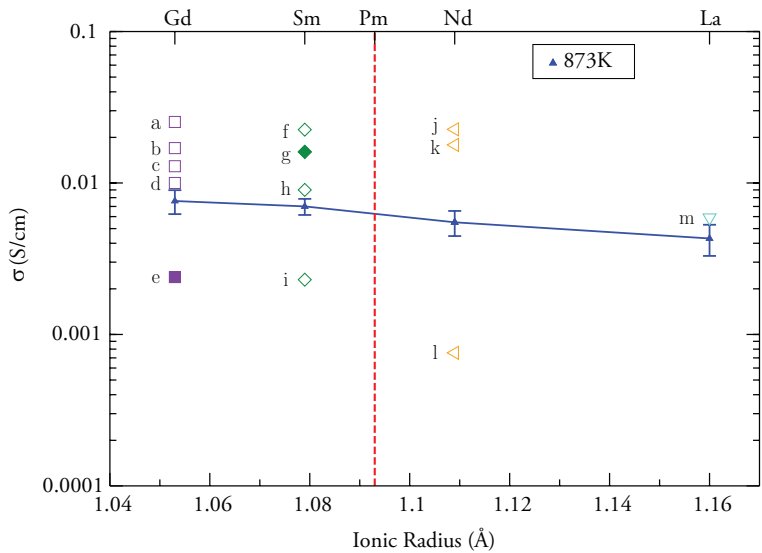}
\end{center}
\caption{Ionic conductivities for Ce$_{0.90}$RE$_{0.10}$O$_{1.95}$ at 873\,K (blue triangles). Open symbols correspond to experimental values, while filled symbols were obtained from computational data in the literature: a) Steele \cite{Steele_SSI2000}, b) Huang \textit{et al.}~\cite{HuangEtAl_JACS1998} and Omar \textit{et al.}~\cite{OmarEtAl_APL2007} c) Xia and Liu~\cite{XiaAndLiu_SSI2002}, d) Zhou \textit{et al.}~\cite{ZhouEtAl_JACS2002}, e) Dholabhai \textit{et al.}~\cite{DholabhaiEtAl_MSMSE2012}, f) Kasse and Nino \cite{KasseAndNino_JAC2013}, g) Grope \textit{et al.} (at 893\,K)~\cite{GropeEtAl_SSI2012} and Omar \textit{et al.}~\cite{OmarEtAl_JACS2009}, h) Shemilt and Williams~\cite{ShemiltAndWilliams_JMSL1999}, i) Jung \textit{et al.}~\cite{JungEtAl_JSSE2002}, j) Kasse and Nino~\cite{KasseAndNino_JAC2013}, k) Omar \textit{et al.}~\cite{OmarEtAl_JACS2009} l) Aneflous \textit{et al.}~\cite{AneflousEtAl_JSSC2004} and m) Dikmen \textit{et al.}~\cite{DikmenEtAl_SSI1999}. The vertical dashed red line represents the critical ionic radius ($r_C$) introduced by Andersson \textit{et al.}~\cite{AnderssonEtAl_PNAS2006}}
\label{Sig873K}
\end{figure*}

Figure \ref{Sig873K} focuses on the DIPPIM simulated ionic conductivities at 873\,K for the systems already presented and puts them in the context of a range of conductivity values from other studies for the same singly doped ceria (Ce$_{0.90}$RE$_{0.10}$O$_{1.95}$) systems at the same temperature. The value for Sc is not included in this plot as there are no data available in the literature for the selected concentration and temperature. Individual literature values are labelled a) to m), with open symbols indicating experimental values and filled symbols those from other computational studies. This figure shows that there is a significant spread in the experimental data set, as exemplified by GdDC whose conductivities vary by a factor of 2.5 (from 0.01\,S/cm \cite{ZhouEtAl_JACS2002} to 0.025\,S/cm \cite{Steele_SSI2000}). Such fluctuations are typically ascribed to different fabrication methods, sintering times and temperatures, grain sizes and impurities \cite{Steele_SSI2000,ZhouEtAl_JACS2002}, all of which are excluded from the our computational bulk models. Nonetheless, what is clear from Figure \ref{Sig873K} is that the DIPPIM potentials used in this work deliver conductivity values which lie in the lower end of the range of the experimental ones, but constitute a good predictor of the overall tendencies in doped ceria. 

\begin{figure*}[htbp]
\begin{center}
\includegraphics[scale=0.4]{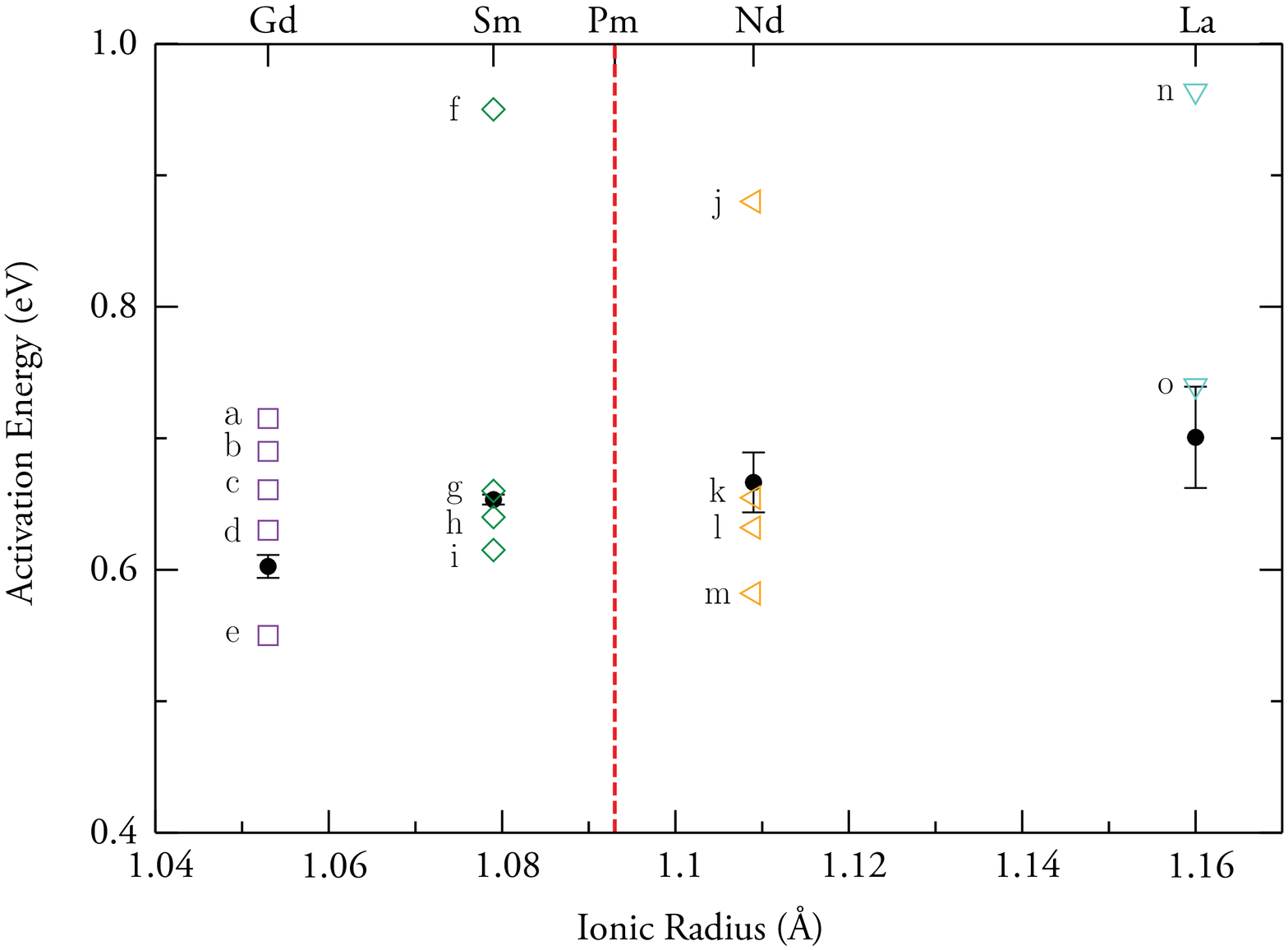}
\end{center}
\caption{Activation energies from this work (filled black circles) between 1473\,K and 873\,K for Ce$_{0.90}$RE$_{0.10}$O$_{1.95}$. Open symbols correspond to experimental values: a) Zhang \textit{et al.}~\cite{ZhangEtAl_SSI2004}, b) Omar \textit{et al.}~\cite{OmarEtAl_APL2007}, c) Omar \textit{et al.}~\cite{OmarEtAl_JACS2009}, d) Huang \textit{et al.}~\cite{HuangEtAl_JACS1998}, e) Xia and Liu~\cite{XiaAndLiu_SSI2002}, f) Jung \textit{et al.}~\cite{JungEtAl_JSSE2002}, g) Shemilt and Williams~\cite{ShemiltAndWilliams_JMSL1999}, h) Omar \textit{et al.}~\cite{OmarEtAl_JACS2009}, i) Kasse and Nino \cite{KasseAndNino_JAC2013}, j) Aneflous \textit{et al.}~\cite{AneflousEtAl_JSSC2004}, k) Omar \textit{et al.}~\cite{OmarEtAl_JACS2009}, l) Kasse and Nino~\cite{KasseAndNino_JAC2013}, m) Stephens and Kilner~\cite{StephensAndKilner_SSI2006}, n) Lang \textit{et al.}~\cite{LangEtAl_JES1999} and o) Dikmen \textit{et al.}~\cite{DikmenEtAl_SSI1999}. The dashed red line represents the critical ionic radius ($r_\mathrm{C}$) introduced by Andersson \textit{et al.}~\cite{AnderssonEtAl_PNAS2006}}
\label{Ea-10pc}
\end{figure*}

The calculated E$_{\mathrm{a}}$ values for these systems are shown in Figure \ref{Ea-10pc} (filled black circles) as a function of dopant ionic radius and were obtained from simulations between 1473\,K and 873\,K. The plots of ln(${\sigma}{T}$) vs 1/$T$ within this range of temperatures were found to be linear for all dopants. This is important as it indicates that there is no extensive clustering of dopants and vacancies, which would occur if the simulations spanned both sides of the critical temperature ($T^{*}$) below which nucleation centres form around the dopants leading to progressive trapping of the vacancies into such clusters as the temperature decreases \cite{HuangEtAl_JACS1998,Goodenough_ARMR2003}. If the calculations had been carried out at sufficiently low temperatures ($T^{*} \approx $ 856\,K for GdDC \cite{HuangEtAl_JACS1998}), then two different E$_{\mathrm{a}}$ values would have been obtained for the $T > T^{*}$ and $T < T^{*}$ temperature regimes. For this reason, the literature data presented in Figure \ref{Ea-10pc} corresponds to the $T > T^{*}$ region only. Experimental values are distinguished with the same open symbols from Figure \ref{Sig873K} and labelled a) to o). Just as was found to be the case with ionic conductivities, Figure \ref{Ea-10pc} illustrates that the self-consistent DIPPIM potentials used in this work are able to predict the activation energies for the various RE dopants used in this study. This is particularly impressive since no experimental data were used at any stage of the potential parameterization. We can therefore proceed to study co-doping in this material, with the confidence that the employed simulation technique can reliably predict the properties of these materials. The predicted E$_{\mathrm{a}}$ for Ce$_{0.90}$Sc$_{0.10}$O$_{1.95}$ is 0.675\,eV. 

\subsection{Co-doped ceria}
\label{Co-doped}

\subsubsection{Lattice constants of co-doped ceria}
\begin{figure*}[htbp]
\begin{center}
\includegraphics[scale=0.4]{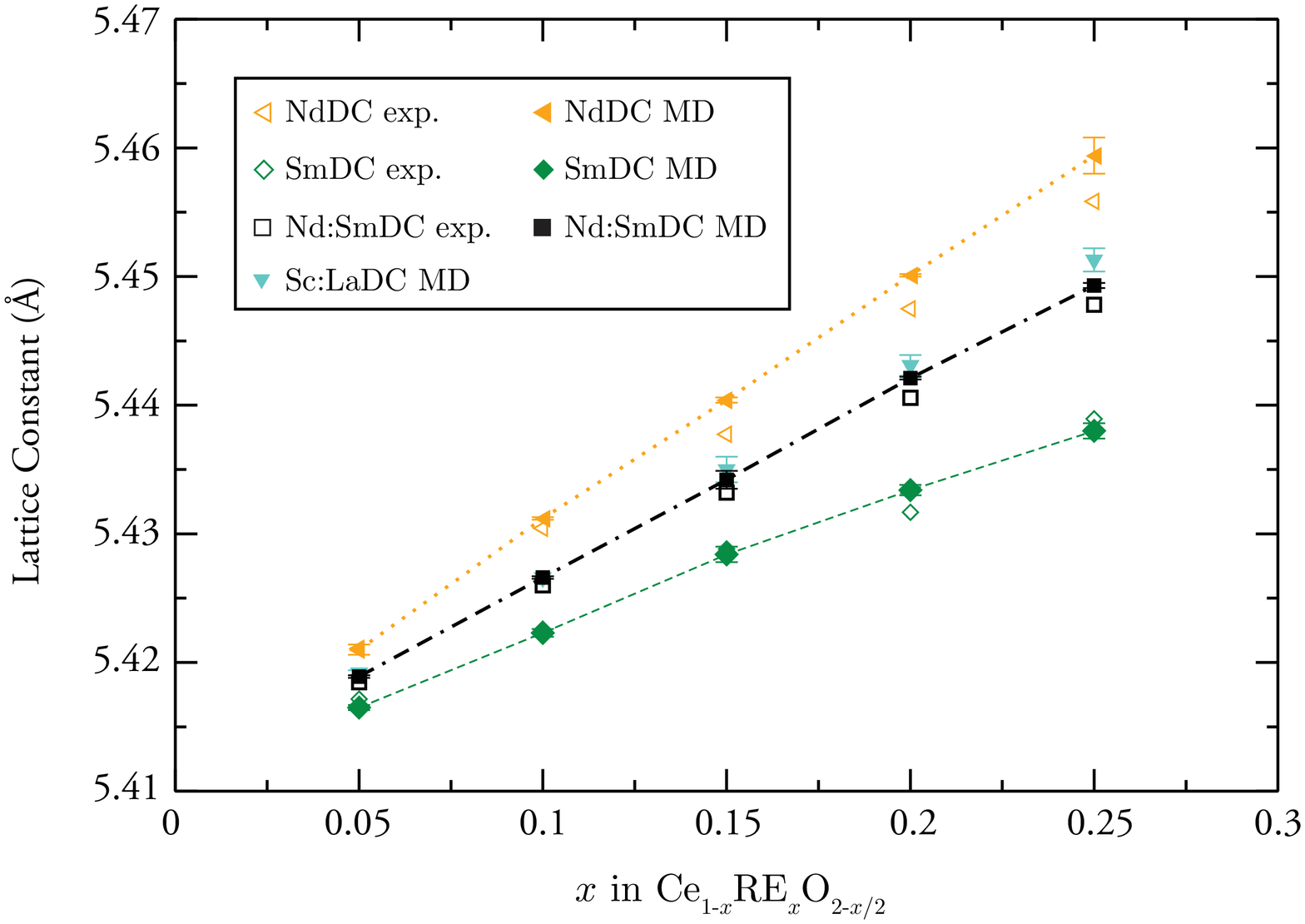}
\end{center}
\caption{DIPPIM calculated (filled symbols) and experimental (open symbols) lattice constants from Buyukkilic \textit{et al.} \cite{BuyukkilicEtAl_SSI2012}at 300\,K for for Ce$_{1-x}$RE$_{x}$O$_{2-x/2}$, where $x$ = 0.05, 0.10, 0.15, 0.20, 0.25 and RE = Sm (diamonds), Nd (triangles pointing left), Nd:SmDC (squares) and Sc:LaDC (triangles pointing down). The lines shown are intended as a guide to the eye only.}
\label{fig:a0-codoped}
\end{figure*}

\ Figure \ref{fig:a0-codoped} presents the DIPPIM calculated (filled symbols) and experimental lattice constants (open symbols) from Buyukkilic \textit{et al.} \cite{BuyukkilicEtAl_SSI2012} at 300\,K for Ce$_{1-x}$RE$_{x}$O$_{2-x/2}$, where $x$ = 0.05, 0.10, 0.15, 0.20, 0.25 and RE = Sm, Nd, Nd:SmDC and Sc:LaDC. The co-doped systems shown correspond to an effective dopant cation radius of 1.093\,\AA\ for the stoichiometries specified in Section \ref{sect:simus}. It is evident from the figure that the calculations predict the correct lattice constant for the singly doped systems over the entire composition range, and that this carries over to the co-doped cerias under study. The agreement is particularly good for the systems that are the focus of this study, namely Ce$_{0.90}$RE$_{0.10}$O$_{1.95}$. The results show that co-doping can be successfully used to reproduce the lattice constant of a cation with a critical dopant radius, $r_\mathrm{C}$. 

\subsubsection{Ionic conductivity of co-doped ceria}\ Thus far the results presented for the ionic conductivity of singly doped ceria have shown only progressive changes in both the bulk ionic conductivities and activation energies as a function of dopant ionic radius. Hence, the question of whether co-doping is a viable alternative for substantially improving these properties becomes: \emph{Do co-doped systems show a marked increase (decrease) in ionic conductivity (activation energy) or is this property simply the average of the singly doped systems?}

\begin{figure*}[htbp]
\begin{center}
\includegraphics[scale=0.32]{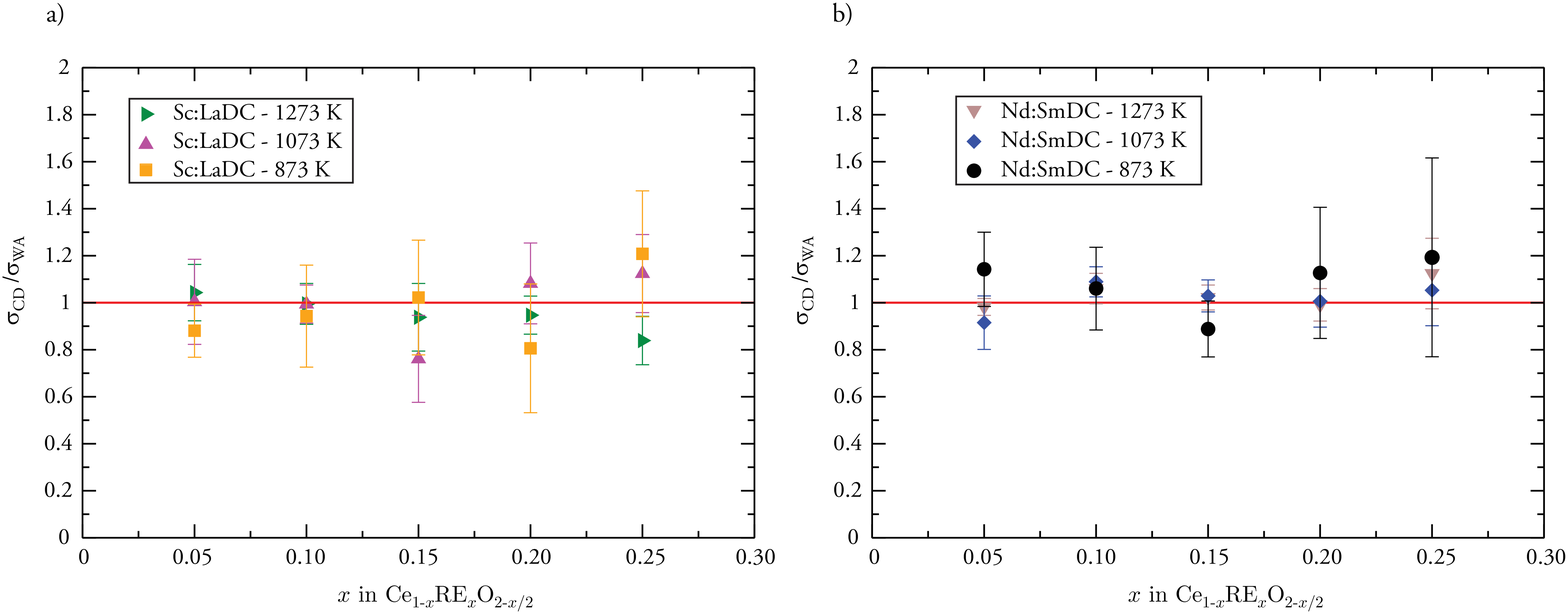}
\end{center}
\caption{DIPPIM ionic conductivities shown as a ratio of the values for co-doped (CD) ceria divided the weighted average (WA) of their singly-doped parent oxides. Panel a) corresponds to the Sc:LaDC system and panel b) corresponds to Nd:SmDC.}
\label{SigAvg}
\end{figure*}

To answer this question it is necessary to compare directly the ionic conductivity of the co-doped systems with that of their ``parent'' singly doped ceria compounds. Hence if there exists a co-doping effect it would be expected that such systems display ionic conductivities that are higher than the average of the values obtained for singly doped ceria. These data are shown in Figure \ref{SigAvg} (a) for the Sc:LaDC system and Figure \ref{SigAvg} (b) for Nd:SmDC. The data in these plots is presented as a ratio of the calculated conductivities for the co-doped systems ($\sigma_{\mathrm{CD}}$) with respect to the weighted average for the singly doped parent compounds with the same total number of dopant cations ($\sigma_{\mathrm{WA}}$). The weighting factors are given by the ratio of each co-dopant as specified in Section \ref{sect:simus}, namely 0.50 for both Nd and Sm in Nd:SmDC, as well as, 0.22 and 0.78 for Sc and La, respectively in Sc:LaDC. The solid red line indicates a linear correspondence between both data sets (co-doped vs weighted average of singly doped), i.e. no co-doping effect. It is clear from these plots that any deviations in $\sigma_{\mathrm{CD}}$ away from the $\sigma_{\mathrm{WA}}$ values are within the margin of error of these measurements (standard deviation). This indicates that the conductivities of co-doped ceria can be predicted by simply calculating the average of the two parent singly doped cerias for all temperatures and dopant concentrations. 

\begin{figure*}[htbp]
\begin{center}
\includegraphics[scale=0.4]{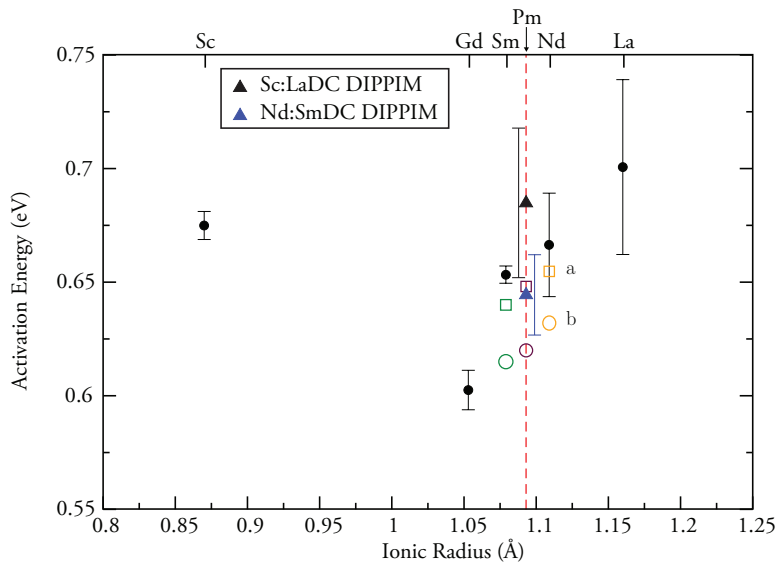}
\end{center}
\caption{DIPPIM activation energies (eV) for Ce$_{0.90}$RE$_{0.10}$O$_{1.95}$ from Figure \ref{Ea-10pc} along with those for the co-doped systems, Nd:SmDC (filled blue triangle) and Sc:LaDC (filled black triangle). Experimental values from a) Omar \textit{et al.} \cite{OmarEtAl_JACS2009} (open squares) and b) Kasse and Nino \cite{KasseAndNino_JAC2013} (open circles) with green for SmDC, maroon for Nd:SmDC and orange for NdDC.}
\label{Ea-10pc-codoped}
\end{figure*}

Accordingly, it is expected that the corresponding activation energies for the ionic conductivity of co-doped ceria display the same averaging effect. This is confirmed in Figure \ref{Ea-10pc-codoped}, which depicts the DIPPIM $E_{a}$ (eV) for Ce$_{0.90}$RE$_{0.10}$O$_{1.95}$ from Figure \ref{Ea-10pc} (filled black circles), along with those for the co-doped systems, Nd:SmDC (filled blue triangle) and Sc:LaDC (filled black triangle). Experimental values from a) Omar \textit{et al.} \cite{OmarEtAl_JACS2009} (open squares) and b) Kasse and Nino \cite{KasseAndNino_JAC2013} (open circles), with green for SmDC, maroon for Nd:SmDC and orange for NdDC. Both data sets show that despite having the same effective $r_\mathrm{C}$ value of 1.093\,\AA and taking into account the errors intrinsic to these calculations, the co-doped Sc:LaDC system has a higher activation energy than Nd:SmDC; that is, the DIPPIM simulations, as well as, the experimental data for these co-doped systems show changes in bulk ionic conductivities and activation energies that are in line with an averaging effect with respect to the singly doped ``parent'' oxides, with small deviations from this behaviour likely due to sampling error. The simple, yet often overlooked, explanation for these patterns is found by analyzing the local structure around the dopants in these systems as shown in the next section.

\subsubsection{Local structure of co-doped systems}
\begin{figure*}[htbp]
\begin{center}
\includegraphics[scale=0.4]{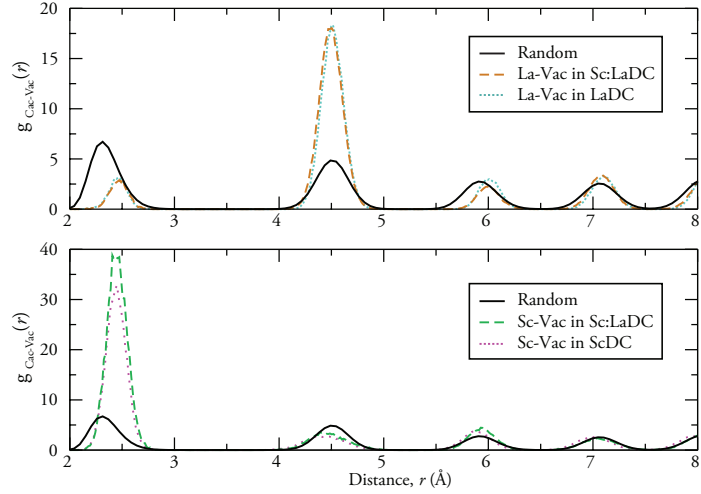}
\end{center}
\caption{Cation--vacancy partial radial distribution functions (g$_{\mathrm{Cat-Vac}}$($r$)) at 873\,K. Top panel: Random (solid black line), La$^{3+}$\,--\,Vacancy in Sc:LaDC (dashed orange line) and La$^{3+}$\,--\,Vac in LaDC (dotted turquoise line). Bottom panel: Random (solid black line), Sc$^{3+}$\,--\,Vacancy in Sc:LaDC (dashed green line) and Sc$^{3+}$\,--\,Vacancy in ScDC (dotted magenta line)}
\label{ScLaDC-CVrdfs}
\end{figure*}

\ A configurational analysis of the MD simulations for 10\% cation doped LaDC, ScDC and Sc:LaDC is illustrated in Figure \ref{ScLaDC-CVrdfs} in the form of the cation-vacancy partial radial distribution functions (g$_{\mathrm{Cat-Vac}}$($r$)) obtained from the average of the three configurations used for each system. Appendix \ref{sect:vacs} details the process of oxygen vacancy identification and subsequent calculation of g($r$). The average values are presented in order to eliminate any possible configuration--dependent ordering of the cations. The La/Sc doubly and singly doped systems are illustrated given that the large radius mismatch between the dopant cations with the host facilitates visualization of the small changes undergone, however the conclusions were confirmed for the other co-doped systems. The top panel shows the g($r$)s for La\,--\,Vac in Sc:LaDC (dashed orange line) vs La\,--\,Vac in LaDC (dotted turquoise line); the bottom panel contains the g($r$)s for Sc\,--\,Vac in Sc:LaDC (dashed green line) vs Sc\,--\,Vac in ScDC (dotted magenta line). The solid black lines in both panels correspond to the Ce\,--\,O g($r$) in bulk ceria at the same temperature, which exemplifies a random vacancy distribution, but with a slightly different lattice constant due to the absence of dopants. The number of vacancies coordinated to the cations in these systems were obtained by integrating the peaks in Figure \ref{ScLaDC-CVrdfs} and are reported in Table \ref{table:CV-integ}, with the addition of the values for GdDC of the same concentration which are included for comparison as it is the best single dopant system. These results show that Sc acts as a vacancy scavenger in ScDC, with the vacancies ordering in the first coordination shell of this cation. This is a well known effect which has been documented by experiments \cite{Gerhardt-AndersonAndNowick_SSI1981,LiEtAl_JACS1991} and simulations \cite{ButlerEtAl_SSI1983,NakayamaAndMartin_PCCP2009}. In fact, dopant cations that are smaller in radius than Ce$^{4+}$ are generally expected to have vacancies in Nearest Neighbour (NN) positions, while those that are larger are expected to have the vacancies in the Next Nearest Neighbour (NNN) position. The latter effect is observed in the case of the larger La and Gd (Table \ref{table:CV-integ}) cations as indicated by the pronounced second peak. The results in Figure \ref{ScLaDC-CVrdfs} and Table \ref{table:CV-integ} clearly show that the local environment, and thus the local strain, of the dopant cations undergoes few changes in going from singly doped systems to those with more than one dopant species.

\begin{table*}[htbp]
\small
\centering
\caption{\ Number of vacancies in the Nearest Neighbour (NN) and Next Nearest Neighbour (NNN) positions with respect to the cations in a random distribution, GdDC, Sc:LaDC, LaDC and ScDC. These values were obtained from the integration of the peaks in Figure \ref{ScLaDC-CVrdfs}}
\label{table:CV-integ}
\begin{tabular}{lllllll}
  \toprule
  \hline
  {Peak} & {Random} & {GdDC} & {LaDC} & {Sc:LaDC} & {ScDC} & {Sc:LaDC} \\
	\cmidrule{2-7}
   & Ce-O  & Gd-Vac & La-Vac & La-Vac & Sc-Vac & Sc-Vac \\
\hline
  1\textsuperscript{st} (NN) & 0.200  & 0.168 & 0.062 & 0.059 & 0.734 & 0.878  \\
  2\textsuperscript{nd} (NNN) & 0.600  & 1.356 & 1.502 & 1.578 & 0.412 & 0.467 \\
  \bottomrule
\end{tabular}
\end{table*}

\begin{figure*}[htbp]
\begin{center}
\includegraphics[scale=0.45]{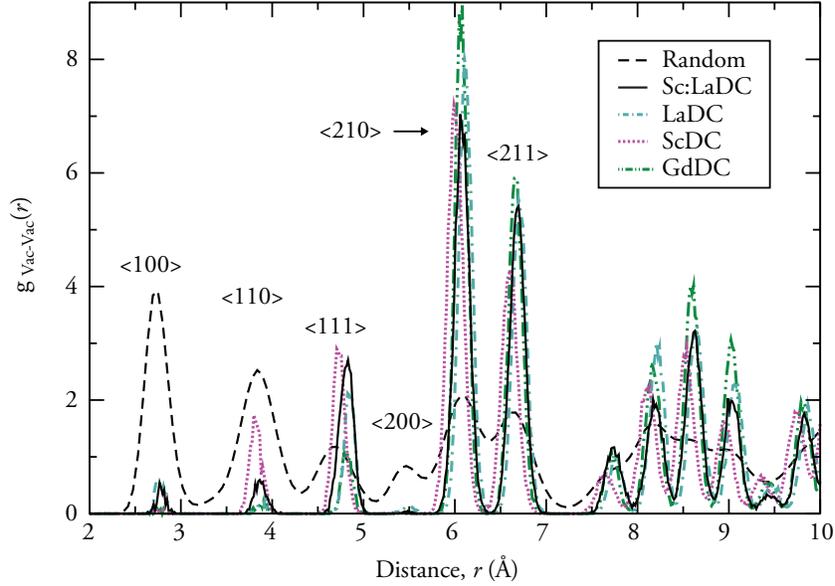}
\end{center}
\caption{Vacancy--vacancy partial radial distribution functions at 873\,K for Sc:LaDC (solid black line), LaDC (dot-dashed turquoise line), ScDC (dotted magenta line), GdDC (dot-dot-dashed green line) and a random vacancy distribution (O--O g($r$)). The $<$100$>$, $<$110$>$ and $<$111$>$ labels indicate different directions along the simple cubic anion sublattice.}
\label{ScLaDC-VVrdfs}
\end{figure*}

Previous studies have shown that concomitant with cation--vacancy ordering, there also exist inherent vacancy--vacancy ordering interactions in fluorite-structured materials \cite{BurbanoEtAl_CM2012,HullEtAl_JSSC2009,MarrocchelliEtAl_CM2011,GopalAndvandeWalle_PRB2012}. Figure \ref{ScLaDC-VVrdfs} presents the three configuration average vacancy--vacancy partial radial distribution functions (g$_{\mathrm{Vac-Vac}}$($r$)) at 873\,K for Sc:LaDC (solid black line), LaDC (dot--dashed turquoise line), ScDC (dotted magenta line), GdDC (dot-dot-dashed green line) as well as a random vacancy distribution, which is simply the O\,--\,O g($r$) from CeO$_{2}$ at the same temperature. The $<$100$>$, $<$110$>$ and $<$111$>$, $<$210$>$, etc labels indicate different directions along the simple cubic anion sublattice. The values obtained upon integration of these peaks are reported in Table \ref{table:VV-integ}. The results show that for this dopant concentration vacancies display some degree of long range ordering as evidenced by the sharp peaks in the $<$210$>$ and $<$211$>$ positions, while the positions that are at shorter distance in the simple cubic lattice are underpopulated with respect to a random vacancy distribution. This effect arises from the Coulomb repulsions between the vacancies. Common to all the doped ceria systems is also a deleterious redistribution of the vacancies which favours short range occupancy along the $<$111$>$ direction with respect to an idealized random system. This effect is larger in inferior conductors, like LaDC and ScDC than in the better ones like GdDC, for example; in the case of Sc:LaDC, co-doping is shown to enhance this ordering, because the small Sc$^{3+}$ cations trap the vacancies in the NN positions, while the large La$^{3+}$ cations repel them towards NNN but also increase their migration barrier \cite{NakayamaAndMartin_PCCP2009}, which leads to an overall increase in the vacancy ordering of co-doped systems. This indicates that the ``synergistic'' effect on bulk ionic conductivity from co-coping is not realized.
\newline

\begin{table}[h]
\small
\centering
\caption{Number of vacancies surrounding another vacancy along the $<$100$>$, $<$110$>$ and $<$111$>$ directions of the simple cubic anion sublattice of 10\% cation doped GdDC, LaDC, ScDC and Sc:LaDC, as well as a random distribution of the same number of vacancies. These values were obtained from the integration of the peaks in Figure \ref{ScLaDC-VVrdfs}}
\label{table:VV-integ}
\begin{tabular}{llll}
  \toprule
  \hline
  System & $<$100$>$ & $<$110$>$ & $<$111$>$ \\
    \midrule
  \hline
  Random & 0.150 & 0.300 & 0.200 \\
	GdDC & 0.001 & 0.010 & 0.056 \\
  LaDC & 0.06 & 0.007 & 0.107 \\
	ScDC & 0.001 & 0.062 & 0.170 \\
%	(LaDC + ScDC)/2 & 0.004 & 0.034 & 0.139 \\
	Sc:LaDC & 0.006 & 0.023 & 0.213 \\
	  \bottomrule
\end{tabular}
\end{table}

\section{Ceria co-doping in perspective}

Despite the significant improvements in the ionic conductivity of co-doped ceria reported by several previous studies \cite{DholabhaiEtAl_JMC2011,OmarEtAl_SSI2006,OmarEtAl_APL2007,Dikmen_JAC2010,DikmenEtAl_JPS2010,ShaEtAl_JAC2007,ShaEtAl_JAC2006,GuanEtAl_MRB2008,WangEtAl_EC2004} this work found that this property is simply an average of the singly doped materials. This is in accordance with an early experimental/computational study by Yoshida \textit{et al.} \cite{YoshidaEtAl_SSI2001,YoshidaEtAl_SSI2003}, as well as, recent experimental data reported Figure \ref{Ea-10pc-codoped} for Nd:SmDC \cite{KasseAndNino_JAC2013,OmarEtAl_JACS2009}. Similar results were reported by Ralph \textit{et al.} for Yb:LaDC and Sm:YDC, and by Li \textit{et al.} \cite{LiEtAl_JACS1991} for Sc:GdDC, who observed a worsening in these properties for co-doped cerias with substantially mismatched dopant cations. The interpretation provided in both cases was that the localized nature of the strains caused by each dopant species does not change substantially in the co-doped systems compared to singly doped materials. \newline

We point out that our investigation has left out a series of factors, such as long-range cation ordering, grain boundaries, impurities, etc, that might also affect the total conductivity of these materials. This was done on purpose, because the focus of this investigation was on bulk properties of perfect fluorite-structured materials. We note here that these factors have usually a detrimental effect on the conductivity of these materials, so that our conclusions are not invalidated by leaving them out. For instance, cation ordering might be expected in Sc:LaDC\cite{GrieshammerEtAl_PCCP2014}, because the cations have significantly different ionic radii. This was not taken into account in these calculations, because the cations were randomly distributed and the simulation timescale does not allow them to diffuse. Cation ordering is known to lead to a decrease of the ionic conductivity, as observed in several oxides \cite{MarrocchelliEtAl_JPCM2009,NorbergEtAl_CM2013}, so that the effects of co-doping might be even more detrimental to the material's conductivity that what we predict in this investigation. \newline

Finally, we wonder why many studies have found an enhancement of the ionic conductivity in some co-doped materials, while others have not. We note that, as discussed above, there are many factors (grain boundaries, cation ordering, phase separation and nano-domain formation, impurity levels, etc) that affect the ionic conductivity of these materials and it is very hard to separate their effects. As an example, the  conductivity of 10\% GdDC, as shown in figure \ref{Sig873K}, varies by as much as a factor of 2.5 in different experiments. Such a huge variation is probably caused by a combination of these factors and shows that it is not trivial to compare the conductivity of these materials.

\section{Conclusions}
\label{Conclusions}

This work was motivated by the conflicting evidence that surrounds the merits of ceria co-doping as a means to improve this electrolyte's ionic conductivity. Here we used Molecular Dynamics simulations that employ accurate interionic potentials, parameterized with respect to {\it first-principles} calculations. No experimental data was used to parameterise these potentials. This methodology allowed the study of large systems at realistic operating temperatures ($\sim$ 873-1273~K) and defect concentrations. The conclusions pertain only to \emph{bulk} properties, because the models that were simulated do not include grain boundaries, impurity segregation, dislocations, etc. \newline

The results show that co-doping can be successfully used to reproduce the lattice constant of ceria doped with a single cation which has an ionic radius equal to the effective radius from two co-dopants. However, close examination of the \textit{bulk} ionic conductivity of co-doped ceria revealed that this property is not enhanced by co-doping and can be described as an average of the conductivities of the ``parent'' singly-doped compounds. This result was explained by the fact that the vacancy ordering tendencies of individual dopant cations remain largely unchanged in co-doped systems. For this reason, co-doping with cations that are bigger/smaller than a given ideal dopant radius ($r_\mathrm{C}$) leads to the combination of unwanted defect trapping tendencies, as exemplified by the case of the significantly mismatched Sc:LaDC system, where Sc$^{3+}$ is an oxygen vacancy scavenger while La$^{3+}$ repels vacancies. In conclusion it is the {\em local} structure of these materials, rather than their average structure, that dictates their conducting properties. These results effectively reject co-doping as a possible avenue for improving ceria conductivity. More fruitful outcomes are likely to be achieved in other intensely investigated areas such as the application of strain \cite{KushimaAndYildizECST2009,KushimaAndYildiz_JMC2010,BurbanoEtAl_JECR2013}.
\newline

\section{Acknowledgements}
MB acknowledges the HPC-EUROPA2 project (project number: 228398) with the support of the European Commission Capacities Area - Research Infrastructures Initiative for access to the computing resources at CINES in France. Calculations were performed on the Stokes supercomputer (allocations tcche026b and tcche031b) maintained by ICHEC and Kelvin maintained by TCHPC. GWW acknowledges Science Foundation Ireland  (PI Grant numbers 12/IA/1414 and 08/RFP/MTR1044). DM wishes to thank the Government of Ireland for an EMPOWER Postdoctoral Fellowship and is thankful to Profs. Harry Tuller and Bilge Yildiz for his recent appointment as visiting scientist at MIT. The authors also thank Prof. Paul A. Madden (Oxford) and Dr Jonathan Yates for helpful discussions.

\appendix

\section{The DIPPIM model} 
\label{DIPPIM}

In this section we provide a brief description of DIPole Polarizable Ionic Model (DIPPIM) potential employed in this work. The reader is referred to \cite{NorbergEtAl_JPCM2009,MaddenEtAl_JMST2006,CastiglioneEtSAl_JPCM1999} and references therein for further information concering the form of the potential. This model has been previously used to study doped ceria and other oxides \cite{MarrocchelliEtAl_JPCM2009,BurbanoEtAl_JPCM2011,BurbanoEtAl_CM2012,Marrocchelli_JPCM2010,WilsonEtAl_JPCM2004}, as well as fluoride systems \cite{HeatonEtAl_JPCB2006,SalanneEtAl_JFC2009}. In DIPPIM the various ionic species that form part of the simulation are assigned their formal valence charges (Ce$^{4+}$, RE$^{3+}$, and O$^{2-}$). The model also includes the polarization effects that result from the induction of dipoles on the ions. The potential is constructed from four components, the first three of which are purely pairwise additive. The DIPPIM components are: charge-charge, dispersion, overlap repulsion and polarization. The Coulombic interactions (charge-charge) are described by:

\begin{equation}
V^{\rm qq}=\sum_{i\leq j}\frac{q_i q_j}{r_{ij}}
\end{equation}

where $q_i$ is the \textit{formal} charge on ion $i$. Dispersion interactions (Equation \ref{dispersion}) include dipole-dipole and dipole-quadrupole terms. Those terms marked in red in Equation \ref{dispersion} and subsequent formulae were part of the fitting process described in Appendix \ref{parameterization}.  

\begin{equation}
\label{dispersion}
V^{\rm disp}=-\sum_{i\leq j }[\frac {f_6^{ij} (r^{ij})  \textcolor{red}{C_6^{ij}}}{r_{ij}^6
}+\frac{f_8^{ij} (r^{ij} ) \textcolor{red}{C_8^{ij}}}{r_{ij}^8 }]
\end{equation}

Here $C_6^{ij}$ and $C_8^{ij}$ are the dipole-dipole and dipole-quadrupole dispersion coefficients, respectively. The $f_n^{ij}$ are Tang-Tonnies damping functions \cite{TangAndToennies_JCP1984,TangAndToennies_JCP2003} which are added in order to describe the short-range penetration correction to the asymptotic dispersion term and are expressed as follows: 

\begin{equation}
\label{tang}
 f_n^{ij}(r_{ij}) = 1 - e^{\textcolor{red}{b^{ij}_{n}} r_{ij}} \sum_{k=0}^n \frac{(\textcolor{red}{b^{ij}_{n}}
 r_{ij})^k}{k!}
\end{equation}

The short range repulsive term (Equation \ref{short-range}) is approximately exponential in the region of physical interionic separations. The full expression used also includes a Gaussian function which acts as a steep repulsive wall and accounts for the anion hard core. This extra term is used in cases where there are highly polarizable ions (e.g. oxygen) to avoid instability problems at very small anion-cation separations \cite{CastiglioneEtSAl_JPCM1999}.

\begin{eqnarray}
\label{short-range}
  V^{\rm rep}= \sum_{i\leq j} \frac{\textcolor{red}{A^{ij}} e^{-\textcolor{red}{a^{ij}} r_{ij}}} {r_{ij}}+ \sum_{i\leq j}\textcolor{red}{B^{ij}} e^{-\textcolor{red}{b^{ij}} r_{ij}^2 }
\end{eqnarray}

As its name indicates it, the polarization part of the DIPPIM potential incorporates dipolar effects only

\begin{eqnarray}
V^\textrm{pol} &=& \sum_{i,j}\left( q_{i}\mu_{j,\alpha} f_4^{ij}(r_{ij})- q_j\mu_{i,\alpha} f_4^{ji}(r_{ij})\right)T_{\alpha}^{(1)}({\bf r}_{ij}) \nonumber \\
& &-\sum_{i,j}\mu_{i,\alpha} \mu_{j,\beta} T_{\alpha\beta}^{(2)}({\bf
r}_{ij})+\sum_i \frac{1}{2\textcolor{red}{\alpha_i}}\mid \boldsymbol{\mu}_i \mid^2
\end{eqnarray}

\noindent Here $\alpha_i$ is the polarizability of ion $i$, $\boldsymbol{\mu}_i$ are the dipoles and ${\bf T}^{(1)}$, ${\bf T}^{(2)}$ are the charge-dipole and dipole-dipole interaction tensors: \\

\begin{equation} 
T_{\alpha}^{(1)}({\bf r})=-r_{\alpha}/r^3 \;\;\;\;\;\;\;\;\; T_{\alpha \beta}^{(2)}({\bf r})=(3r_{\alpha} r_{\beta}-r^2\delta_{\alpha \beta})/r^5
\end{equation}

The instantaneous values of the dipole moments are obtained by minimization of this expression with respect to the dipoles of all ions at each MD timestep. This ensures that we regain the condition that the dipole induced by an electrical field ${\bf E}$ is $\alpha {\bf E}$ and that the dipole values are mutually consistent. The short-range induction effects on the dipoles are taken into account by the Tang-Toennies damping functions ($f_4^{ij}$) similar to those used to damp the dispersion interactions (Equation \ref{tang}) where $b^{ij}$ determines the range at which the overlap of the charge densities affects the induced dipoles. The damping of these induction effects requires an additional pre-exponential parameter, $c^{ij}$, which determines the strength of the ion response to this effect.

\section{Potential parameterization}
\label{parameterization}

A total of 19 2 x 2 x 2 fluorite-structured supercells were used to fit this IP set. They included YSZ (1), ScSZ (1), ceria-zirconia (3), pure ceria (1), reduced ceria (1) and two configurations of composition Ce$_{0.5}$RE$_{0.5}$O$_{1.75}$ for each of the systems. Each model supercell was obtained from high temperature (2000\,K) \textit{ab initio} MD simulations that were run for a few pico seconds in order to reach structural equilibrium. The forces on each species were determined directly from each DFT calculation, and the dipoles were obtained from a Wannier analysis of the Kohn-Sham (KS) wave functions \cite{MarzariAndVanderbilt_PRB1997}. In each case, hybrid density functional theory (h-DFT) calculations using the Heyd, Scuseria, Ernzerhof (HSE06) functional \cite{HeydEtAl_JCP2003,KrukauEtAl_JCP2006}, as implemented in the VASP code \cite{PaierEtAl_JCP2006} were performed. The inclusion of a fraction of nonlocal Hartree-Fock exchange to standard DFT in functionals such as HSE is known to be necessary to correctly describe the electronic strucuture of lanthanide oxides, such as, reduced ceria \cite{DaSilvaEtAl_PRB2007,GillenEtAl_PRB2013} due to the highly correlated \textit{f}-electrons present in these elements. This DFT functional also represents an improvement over other more commonly used functionals like LDA (Local Density Approximation) or GGA (Generalized Gradient Approximation) in terms of a closer agreement to experimental lattice constants \cite{DaSilvaEtAl_PRB2007} in cases where there are no $f$-electrons present, e.g. YDC, LaDC, etc.  
 
Although dispersion energies constitute only a small fraction of the total energy, they have a considerable influence on transition pressures and, in particular, on the material density, thus the lattice constant, and the stress tensor. However, the dispersion terms were not included in the initial fit due to the well known uncontrolled representation of dispersion within the framework of DFT \cite{KochEtAndHolthausen_Book2001}. Instead, values for the C$_{6}^{ij}$ and C$_{8}^{ij}$ terms were determined from the dipole polarizabilities for each element that resulted from the initial potential fit. The $\alpha$ values thus obtained have been previously found to be in very good agreement with other theoretically derived values \cite{BurbanoEtAl_JPCM2011,HeatonEtAl_JCP2006}. The relationship between the dispersion coefficients and the dipole polarizability is given by the Slater-Kirkwood equation \cite{SlaterAndKirkwood_PR1931,HaigisEtAl_CG2013}. The final values for the DIPPIM parameters were then obtained by fixing the values of the polarizabilities and dispersion terms in a last round of optimization and re-fitting. The final $\chi^{2}$ values for fit were 0.216 for dipoles and 0.335 for forces. Table \ref{DIPPIMParams} presents the parameters for the DIPPIM interionic potential set that was obtained.
\newline

\begin{table*}[h]
\centering
\small
\caption{Parameters for the DIPPIM potential. All values are in atomic units, except those corresponding to the ionic radii, which are given in angstroms (\AA) and are shown in parentheses alongside the dipole polarizabilities $\alpha$. The parameters $b_D^{\rm{O}^{2-} \;-\; \square}$ and $b_D^{\square \;-\; \rm{O}^{2-}}$ were given the same value. Here, $\square$ represents a placeholder for the identity of the ionic species specified in a given column}
\label{DIPPIMParams}
\begin{tabular}{c c c c c c c c c}
\toprule
\hline
  Interaction       	      & $A^{ij}$ & $a^{ij}$ & $B^{ij}$ & $b^{ij}$ & $C_6^{ij}$ & $C_8^{ij}$ & $b_{6}^{ij}$ & $b_{8}^{ij}$ \\
\midrule
\hline
O$^{2-}$ -- O$^{2-}$ &  7.15 & 18.52 & 50000 & 1.00 & 83 & 1240 & 1.30 & 1.70 \\
Ce$^{4+}$ -- O$^{2-}$ & 82.20	& 1.19 	& 50000 & 1.55	& 47 & 595 & 1.50 & 1.96 \\
Zr$^{4+}$ -- O$^{2-}$ & 89.79	& 1.29 	& 50000 & 1.75	& 21 & 271 & 1.62 & 2.10 \\
La$^{3+}$ -- O$^{2-}$ & 102.63 	& 1.25 & 50000 & 1.30 & 57 & 731 & 1.46 & 1.88 \\
Ce$^{3+}$ -- O$^{2-}$ & 100.02 	& 1.25 & 50000 & 1.20 & 71 & 902 & 1.47 & 1.90 \\
Nd$^{3+}$ -- O$^{2-}$ & 94.24	& 1.25 & 50000 & 1.36 & 56 & 709 & 1.49 & 1.94 \\
Sm$^{3+}$ -- O$^{2-}$ & 87.79 & 1.25 & 50000 & 1.38 & 49 & 630 & 1.51 & 1.97 \\
Gd$^{3+}$ -- O$^{2-}$ & 79.98	& 1.25 & 50000 & 1.38 & 23 & 293 & 1.54&  2.00 \\
Y$^{3+}$ -- O$^{2-}$ & 118.0 	& 1.38 	& 50000 &	1.50 & 21 & 264 & 1.60 & 2.08 \\
Sc$^{3+}$ -- O$^{2-}$ & 61.66	& 1.28 & 50000 & 1.75 & 15 & 197 & 1.77 & 2.30 \\
\midrule
\hline
Ion &$\alpha$ & radius (\AA) & $b_D^{\rm{O}^{2-} \;-\; \square}$ & $c_D^{\rm{O}^{2-} \;-\; \square}$  & $c_D^{\square \;-\; \rm{O}^{2-}}$ \\
\midrule
\hline
O$^{2-}$		& 13.97 & 1.38 & 2.18 & 3.03 & -- \\
Ce$^{4+}$		& 5.86 & 0.97 & 1.75 	& 1.85 & 0.17 \\
Zr$^{4+}$		& 2.38 & 0.84 & 1.74 	& 1.56 & -0.60 \\
La$^{3+}$		& 7.51 & 1.16 & 1.50 	& 1.43 & -0.20 \\
Ce$^{3+}$		& 9.72 & 1.143& 1.59 	& 1.71 & 0.05 \\
Nd$^{3+}$		& 7.24 & 1.109& 1.67 	& 1.94 & 0.00 \\
Sm$^{3+}$		& 6.28 & 1.079& 1.67 	& 1.97 & -0.14 \\
Gd$^{3+}$		& 2.56 & 1.053& 1.69  & 1.75 & -0.89 \\
Y$^{3+}$		& 2.31 & 1.019& 1.47 	& 1.08 & -0.60 \\
Sc$^{3+}$		& 1.70 & 0.87 & 1.67 	& 1.36 & -0.39 \\
\bottomrule
\end{tabular}
\end{table*}
\section{Vacancy analysis} 
\label{sect:vacs}

The vacancy analysis of fluorite structured materials probes the occupancy changes of the tetrahedral sites formed by the face centred cubic cation sublattice. These are the sites where the oxide ions and, thus, their vacancies are normally present. The oxygen vacancy analysis is performed making use of the fact that the cations in the systems under study do not diffuse, even at high temperatures. Thus, it is possible to specify the cation sublattice in terms of tetrahedra, each of which may be empty (vacancy) or occupied by an oxide anion. For a set of time-correlated instantaneous ionic configurations, i.e., frames in an MD simulation, one can determine which tetrahedral sites are empty. However, distinction must be made between the vibrations undergone by the oxide anions, which are large in amplitude at the temperatures of interest, and instances when a given oxide anion has truly vacated a given tetrahedron. This distinction is made by imposing the condition that a site be considered a vacancy only if it has been vacant for a period of at least two frames. The position of each vacancy is then defined as the centre of the tetrahedron formed by the average positions of the four surrounding cations. These positions can then be used to calculate partial radial distribution functions (RDFs), from which vacancy ordering in real space can be studied. Integration of the vacancy-vacancy RDF ($g_{v-v}$) peaks from zero out to the position $r_{c}$ gives the vacancy-vacancy coordination number, $n_{v-v}$: 

\begin{equation}
n_{v-v} = 4\pi\rho \int_{0}^{r_{c}}{r^{2}g_{v-v}dr}
\label{eq:vac_gr}
\end{equation}

\noindent where $\rho$ is the vacancy density in the simulation cell. This method has been previously used to study similar materials, such as, Zr$_{2}$Y$_{2}$O$_{7}$--Y$_{3}$NbO$_{7}$ \cite{MarrocchelliEtAl_JPCM2009}, YDC \cite{BurbanoEtAl_CM2012} and PbO$_{2}$ \cite{CastiglioneEtSAl_JPCM1999,CastiglioneAndMadden_JPCM2001}.


%\balance


%\footnotesize{
%\bibliography{BIB} %your .bib file
%\bibliographystyle{rsc} %the RSC's .bst file
%}
%\bibliography{BIB}

\begin{thebibliography}{122}%
\makeatletter
\providecommand \@ifxundefined [1]{%
 \@ifx{#1\undefined}
}%
\providecommand \@ifnum [1]{%
 \ifnum #1\expandafter \@firstoftwo
 \else \expandafter \@secondoftwo
 \fi
}%
\providecommand \@ifx [1]{%
 \ifx #1\expandafter \@firstoftwo
 \else \expandafter \@secondoftwo
 \fi
}%
\providecommand \natexlab [1]{#1}%
\providecommand \enquote  [1]{``#1''}%
\providecommand \bibnamefont  [1]{#1}%
\providecommand \bibfnamefont [1]{#1}%
\providecommand \citenamefont [1]{#1}%
\providecommand \href@noop [0]{\@secondoftwo}%
\providecommand \href [0]{\begingroup \@sanitize@url \@href}%
\providecommand \@href[1]{\@@startlink{#1}\@@href}%
\providecommand \@@href[1]{\endgroup#1\@@endlink}%
\providecommand \@sanitize@url [0]{\catcode `\\12\catcode `\$12\catcode
  `\&12\catcode `\#12\catcode `\^12\catcode `\_12\catcode `\%12\relax}%
\providecommand \@@startlink[1]{}%
\providecommand \@@endlink[0]{}%
\providecommand \url  [0]{\begingroup\@sanitize@url \@url }%
\providecommand \@url [1]{\endgroup\@href {#1}{\urlprefix }}%
\providecommand \urlprefix  [0]{URL }%
\providecommand \Eprint [0]{\href }%
\providecommand \doibase [0]{http://dx.doi.org/}%
\providecommand \selectlanguage [0]{\@gobble}%
\providecommand \bibinfo  [0]{\@secondoftwo}%
\providecommand \bibfield  [0]{\@secondoftwo}%
\providecommand \translation [1]{[#1]}%
\providecommand \BibitemOpen [0]{}%
\providecommand \bibitemStop [0]{}%
\providecommand \bibitemNoStop [0]{.\EOS\space}%
\providecommand \EOS [0]{\spacefactor3000\relax}%
\providecommand \BibitemShut  [1]{\csname bibitem#1\endcsname}%
\let\auto@bib@innerbib\@empty
%</preamble>
\bibitem [{\citenamefont {Karger}\ and\ \citenamefont
  {Hennings}(2009)}]{KargerAndHennings_RSER2009}%
  \BibitemOpen
  \bibfield  {author} {\bibinfo {author} {\bibfnamefont {C.~R.}\ \bibnamefont
  {Karger}}\ and\ \bibinfo {author} {\bibfnamefont {W.}~\bibnamefont
  {Hennings}},\ }\href@noop {} {\bibfield  {journal} {\bibinfo  {journal}
  {Renewable and Sustainable Energy Reviews}\ }\textbf {\bibinfo {volume}
  {13}},\ \bibinfo {pages} {583 } (\bibinfo {year} {2009})}\BibitemShut
  {NoStop}%
\bibitem [{\citenamefont {IEA}(2002)}]{DE_IEA2002}%
  \BibitemOpen
  \bibfield  {author} {\bibinfo {author} {\bibnamefont {IEA}},\ }\href@noop {}
  {\emph {\bibinfo {title} {Distributed Generation in Liberalised Electricity
  Markets}}}\ (\bibinfo  {publisher} {OECD Publishing (Paris)},\ \bibinfo
  {year} {2002})\ pp.~\bibinfo {pages} {--}\BibitemShut {NoStop}%
\bibitem [{\citenamefont {IEA}(2008)}]{CHP_IEA2008}%
  \BibitemOpen
  \bibfield  {author} {\bibinfo {author} {\bibnamefont {IEA}},\ }\href@noop {}
  {\emph {\bibinfo {title} {Combined Heat and Power}}}\ (\bibinfo  {publisher}
  {OECD Publishing (Paris)},\ \bibinfo {year} {2008})\BibitemShut {NoStop}%
\bibitem [{\citenamefont {Ormerod}(2003)}]{Ormerod_CSR2003}%
  \BibitemOpen
  \bibfield  {author} {\bibinfo {author} {\bibfnamefont {R.~M.}\ \bibnamefont
  {Ormerod}},\ }\href@noop {} {\bibfield  {journal} {\bibinfo  {journal} {Chem
  Soc Rev}\ }\textbf {\bibinfo {volume} {32}},\ \bibinfo {pages} {17 }
  (\bibinfo {year} {2003})}\BibitemShut {NoStop}%
\bibitem [{\citenamefont {Guo}\ \emph {et~al.}(2013)\citenamefont {Guo},
  \citenamefont {Bessaa}, \citenamefont {Aguado}, \citenamefont {Steil},
  \citenamefont {Rembelski}, \citenamefont {Rieu}, \citenamefont {Viricelle},
  \citenamefont {Benameur}, \citenamefont {Guizard}, \citenamefont {Tardivat},
  \citenamefont {Vernoux},\ and\ \citenamefont {Farrusseng}}]{GuoEtAl_EES2013}%
  \BibitemOpen
  \bibfield  {author} {\bibinfo {author} {\bibfnamefont {Y.}~\bibnamefont
  {Guo}}, \bibinfo {author} {\bibfnamefont {M.}~\bibnamefont {Bessaa}},
  \bibinfo {author} {\bibfnamefont {S.}~\bibnamefont {Aguado}}, \bibinfo
  {author} {\bibfnamefont {M.~C.}\ \bibnamefont {Steil}}, \bibinfo {author}
  {\bibfnamefont {D.}~\bibnamefont {Rembelski}}, \bibinfo {author}
  {\bibfnamefont {M.}~\bibnamefont {Rieu}}, \bibinfo {author} {\bibfnamefont
  {J.-P.}\ \bibnamefont {Viricelle}}, \bibinfo {author} {\bibfnamefont
  {N.}~\bibnamefont {Benameur}}, \bibinfo {author} {\bibfnamefont
  {C.}~\bibnamefont {Guizard}}, \bibinfo {author} {\bibfnamefont
  {C.}~\bibnamefont {Tardivat}}, \bibinfo {author} {\bibfnamefont
  {P.}~\bibnamefont {Vernoux}}, \ and\ \bibinfo {author} {\bibfnamefont
  {D.}~\bibnamefont {Farrusseng}},\ }\href@noop {} {\bibfield  {journal}
  {\bibinfo  {journal} {Energy Environ. Sci.}\ }\textbf {\bibinfo {volume}
  {6}},\ \bibinfo {pages} {2119} (\bibinfo {year} {2013})}\BibitemShut
  {NoStop}%
\bibitem [{\citenamefont {Wachsman}\ \emph {et~al.}(2012)\citenamefont
  {Wachsman}, \citenamefont {Marlowe},\ and\ \citenamefont
  {Lee}}]{WachsmanEtAl_EES2012}%
  \BibitemOpen
  \bibfield  {author} {\bibinfo {author} {\bibfnamefont {E.~D.}\ \bibnamefont
  {Wachsman}}, \bibinfo {author} {\bibfnamefont {C.~A.}\ \bibnamefont
  {Marlowe}}, \ and\ \bibinfo {author} {\bibfnamefont {K.~T.}\ \bibnamefont
  {Lee}},\ }\href@noop {} {\bibfield  {journal} {\bibinfo  {journal} {Energy
  Environ. Sci.}\ }\textbf {\bibinfo {volume} {5}},\ \bibinfo {pages} {5498}
  (\bibinfo {year} {2012})}\BibitemShut {NoStop}%
\bibitem [{\citenamefont {Wachsman}\ and\ \citenamefont
  {Lee}(2011)}]{WachsmanAndLee_S2011}%
  \BibitemOpen
  \bibfield  {author} {\bibinfo {author} {\bibfnamefont {E.~D.}\ \bibnamefont
  {Wachsman}}\ and\ \bibinfo {author} {\bibfnamefont {K.~T.}\ \bibnamefont
  {Lee}},\ }\href@noop {} {\bibfield  {journal} {\bibinfo  {journal} {Science}\
  }\textbf {\bibinfo {volume} {334}},\ \bibinfo {pages} {935} (\bibinfo {year}
  {2011})}\BibitemShut {NoStop}%
\bibitem [{\citenamefont {Qin}\ \emph {et~al.}(2011)\citenamefont {Qin},
  \citenamefont {Zhu}, \citenamefont {Liu}, \citenamefont {Jing}, \citenamefont
  {Raza}, \citenamefont {Imran}, \citenamefont {Singh}, \citenamefont {Abbas},\
  and\ \citenamefont {Zhu}}]{QinEtAl_EES2011}%
  \BibitemOpen
  \bibfield  {author} {\bibinfo {author} {\bibfnamefont {H.}~\bibnamefont
  {Qin}}, \bibinfo {author} {\bibfnamefont {Z.}~\bibnamefont {Zhu}}, \bibinfo
  {author} {\bibfnamefont {Q.}~\bibnamefont {Liu}}, \bibinfo {author}
  {\bibfnamefont {Y.}~\bibnamefont {Jing}}, \bibinfo {author} {\bibfnamefont
  {R.}~\bibnamefont {Raza}}, \bibinfo {author} {\bibfnamefont {S.}~\bibnamefont
  {Imran}}, \bibinfo {author} {\bibfnamefont {M.}~\bibnamefont {Singh}},
  \bibinfo {author} {\bibfnamefont {G.}~\bibnamefont {Abbas}}, \ and\ \bibinfo
  {author} {\bibfnamefont {B.}~\bibnamefont {Zhu}},\ }\href@noop {} {\bibfield
  {journal} {\bibinfo  {journal} {Energy Environ. Sci.}\ }\textbf {\bibinfo
  {volume} {4}},\ \bibinfo {pages} {1273} (\bibinfo {year} {2011})}\BibitemShut
  {NoStop}%
\bibitem [{\citenamefont {Ruiz-Morales}\ \emph {et~al.}(2010)\citenamefont
  {Ruiz-Morales}, \citenamefont {Marrero-Lopez}, \citenamefont
  {Galvez-Sanchez}, \citenamefont {Canales-Vazquez}, \citenamefont {Savaniu},\
  and\ \citenamefont {Savvin}}]{Ruiz-MoralesEtAl_ESS2010}%
  \BibitemOpen
  \bibfield  {author} {\bibinfo {author} {\bibfnamefont {J.~C.}\ \bibnamefont
  {Ruiz-Morales}}, \bibinfo {author} {\bibfnamefont {D.}~\bibnamefont
  {Marrero-Lopez}}, \bibinfo {author} {\bibfnamefont {M.}~\bibnamefont
  {Galvez-Sanchez}}, \bibinfo {author} {\bibfnamefont {J.}~\bibnamefont
  {Canales-Vazquez}}, \bibinfo {author} {\bibfnamefont {C.}~\bibnamefont
  {Savaniu}}, \ and\ \bibinfo {author} {\bibfnamefont {S.~N.}\ \bibnamefont
  {Savvin}},\ }\href {\doibase 10.1039/C0EE00166J} {\bibfield  {journal}
  {\bibinfo  {journal} {Energy Environ. Sci.}\ }\textbf {\bibinfo {volume}
  {3}},\ \bibinfo {pages} {1670} (\bibinfo {year} {2010})}\BibitemShut
  {NoStop}%
\bibitem [{\citenamefont {Brett}\ \emph {et~al.}(2008)\citenamefont {Brett},
  \citenamefont {Atkinson}, \citenamefont {Brandon},\ and\ \citenamefont
  {Skinner}}]{BrettEtAl_CSR2008}%
  \BibitemOpen
  \bibfield  {author} {\bibinfo {author} {\bibfnamefont {D.~J.~L.}\
  \bibnamefont {Brett}}, \bibinfo {author} {\bibfnamefont {A.}~\bibnamefont
  {Atkinson}}, \bibinfo {author} {\bibfnamefont {N.~P.}\ \bibnamefont
  {Brandon}}, \ and\ \bibinfo {author} {\bibfnamefont {S.~J.}\ \bibnamefont
  {Skinner}},\ }\href@noop {} {\bibfield  {journal} {\bibinfo  {journal} {Chem
  Soc Rev}\ }\textbf {\bibinfo {volume} {37}},\ \bibinfo {pages} {1568}
  (\bibinfo {year} {2008})}\BibitemShut {NoStop}%
\bibitem [{\citenamefont {Tuller}\ and\ \citenamefont
  {Nowick}(1975)}]{TullerAndNowick_JES1975}%
  \BibitemOpen
  \bibfield  {author} {\bibinfo {author} {\bibfnamefont {H.~L.}\ \bibnamefont
  {Tuller}}\ and\ \bibinfo {author} {\bibfnamefont {A.~S.}\ \bibnamefont
  {Nowick}},\ }\href@noop {} {\bibfield  {journal} {\bibinfo  {journal} {J
  Electrochem Soc}\ }\textbf {\bibinfo {volume} {122}},\ \bibinfo {pages} {255}
  (\bibinfo {year} {1975})}\BibitemShut {NoStop}%
\bibitem [{\citenamefont {Tuller}\ and\ \citenamefont
  {Nowick}(1977)}]{TullerAndNowick_JPCS1977}%
  \BibitemOpen
  \bibfield  {author} {\bibinfo {author} {\bibfnamefont {H.~L.}\ \bibnamefont
  {Tuller}}\ and\ \bibinfo {author} {\bibfnamefont {A.~S.}\ \bibnamefont
  {Nowick}},\ }\href@noop {} {\bibfield  {journal} {\bibinfo  {journal} {J Phys
  Chem Solids}\ }\textbf {\bibinfo {volume} {38}},\ \bibinfo {pages} {859 }
  (\bibinfo {year} {1977})}\BibitemShut {NoStop}%
\bibitem [{Note1()}]{Note1}%
  \BibitemOpen
  \bibinfo {note} {$\delta $-Bi$_2$O$_3$ actually behaves differently, since
  this material already has 25\% intrinsic vacancies. In this case doping with
  iso-valent cations is used to stabilise the fluorite structure at low
  temperatures \cite {AbrahamsEtAL_CM2010}.}\BibitemShut {Stop}%
\bibitem [{\citenamefont {Sun}\ \emph {et~al.}(2012)\citenamefont {Sun},
  \citenamefont {Li},\ and\ \citenamefont {Chen}}]{SunEtAl_EES2012}%
  \BibitemOpen
  \bibfield  {author} {\bibinfo {author} {\bibfnamefont {C.}~\bibnamefont
  {Sun}}, \bibinfo {author} {\bibfnamefont {H.}~\bibnamefont {Li}}, \ and\
  \bibinfo {author} {\bibfnamefont {L.}~\bibnamefont {Chen}},\ }\href@noop {}
  {\bibfield  {journal} {\bibinfo  {journal} {Energy Environ. Sci.}\ }\textbf
  {\bibinfo {volume} {5}},\ \bibinfo {pages} {8475} (\bibinfo {year}
  {2012})}\BibitemShut {NoStop}%
\bibitem [{\citenamefont {Bogicevic}\ \emph {et~al.}(2001)\citenamefont
  {Bogicevic}, \citenamefont {Wolverton}, \citenamefont {Crosbie},\ and\
  \citenamefont {Stechel}}]{BogicevicEtAl_PRB2001}%
  \BibitemOpen
  \bibfield  {author} {\bibinfo {author} {\bibfnamefont {A.}~\bibnamefont
  {Bogicevic}}, \bibinfo {author} {\bibfnamefont {C.}~\bibnamefont
  {Wolverton}}, \bibinfo {author} {\bibfnamefont {G.~M.}\ \bibnamefont
  {Crosbie}}, \ and\ \bibinfo {author} {\bibfnamefont {E.~B.}\ \bibnamefont
  {Stechel}},\ }\href@noop {} {\bibfield  {journal} {\bibinfo  {journal}
  {Physical Review B}\ }\textbf {\bibinfo {volume} {64}},\ \bibinfo {pages}
  {014106} (\bibinfo {year} {2001})}\BibitemShut {NoStop}%
\bibitem [{\citenamefont {Bogicevic}\ and\ \citenamefont
  {Wolverton}(2003)}]{BogicevicAndWolverton_PRB2003}%
  \BibitemOpen
  \bibfield  {author} {\bibinfo {author} {\bibfnamefont {A.}~\bibnamefont
  {Bogicevic}}\ and\ \bibinfo {author} {\bibfnamefont {C.}~\bibnamefont
  {Wolverton}},\ }\href@noop {} {\bibfield  {journal} {\bibinfo  {journal}
  {Physical Review B}\ }\textbf {\bibinfo {volume} {67}},\ \bibinfo {pages}
  {024106} (\bibinfo {year} {2003})}\BibitemShut {NoStop}%
\bibitem [{\citenamefont {Navrotsky}\ \emph {et~al.}(2007)\citenamefont
  {Navrotsky}, \citenamefont {Simoncic}, \citenamefont {Yokokawa},
  \citenamefont {Chen},\ and\ \citenamefont {Lee}}]{NavrotskyEtAl_FD2007}%
  \BibitemOpen
  \bibfield  {author} {\bibinfo {author} {\bibfnamefont {A.}~\bibnamefont
  {Navrotsky}}, \bibinfo {author} {\bibfnamefont {P.}~\bibnamefont {Simoncic}},
  \bibinfo {author} {\bibfnamefont {H.}~\bibnamefont {Yokokawa}}, \bibinfo
  {author} {\bibfnamefont {W.}~\bibnamefont {Chen}}, \ and\ \bibinfo {author}
  {\bibfnamefont {T.}~\bibnamefont {Lee}},\ }\href@noop {} {\bibfield
  {journal} {\bibinfo  {journal} {Faraday Discuss}\ }\textbf {\bibinfo {volume}
  {134}},\ \bibinfo {pages} {171} (\bibinfo {year} {2007})}\BibitemShut
  {NoStop}%
\bibitem [{\citenamefont {Pietrucci}\ \emph {et~al.}(2008)\citenamefont
  {Pietrucci}, \citenamefont {Bernasconi}, \citenamefont {Laio},\ and\
  \citenamefont {Parrinello}}]{PietrucciEtAl_PRB2008}%
  \BibitemOpen
  \bibfield  {author} {\bibinfo {author} {\bibfnamefont {F.}~\bibnamefont
  {Pietrucci}}, \bibinfo {author} {\bibfnamefont {M.}~\bibnamefont
  {Bernasconi}}, \bibinfo {author} {\bibfnamefont {A.}~\bibnamefont {Laio}}, \
  and\ \bibinfo {author} {\bibfnamefont {M.}~\bibnamefont {Parrinello}},\
  }\href@noop {} {\bibfield  {journal} {\bibinfo  {journal} {Physical Review
  B}\ }\textbf {\bibinfo {volume} {78}},\ \bibinfo {pages} {094301} (\bibinfo
  {year} {2008})}\BibitemShut {NoStop}%
\bibitem [{\citenamefont {Navrotsky}(2010)}]{Navrotsky_JMC2010}%
  \BibitemOpen
  \bibfield  {author} {\bibinfo {author} {\bibfnamefont {A.}~\bibnamefont
  {Navrotsky}},\ }\href@noop {} {\bibfield  {journal} {\bibinfo  {journal} {J
  Mater Chem}\ }\textbf {\bibinfo {volume} {20}},\ \bibinfo {pages} {10577}
  (\bibinfo {year} {2010})}\BibitemShut {NoStop}%
\bibitem [{\citenamefont {Norberg}\ \emph {et~al.}(2011)\citenamefont
  {Norberg}, \citenamefont {Hull}, \citenamefont {Ahmed}, \citenamefont
  {Eriksson}, \citenamefont {Marrocchelli}, \citenamefont {Madden},
  \citenamefont {Li},\ and\ \citenamefont {Irvine}}]{NorbergEtAl_CM2011}%
  \BibitemOpen
  \bibfield  {author} {\bibinfo {author} {\bibfnamefont {S.~T.}\ \bibnamefont
  {Norberg}}, \bibinfo {author} {\bibfnamefont {S.}~\bibnamefont {Hull}},
  \bibinfo {author} {\bibfnamefont {I.}~\bibnamefont {Ahmed}}, \bibinfo
  {author} {\bibfnamefont {S.~G.}\ \bibnamefont {Eriksson}}, \bibinfo {author}
  {\bibfnamefont {D.}~\bibnamefont {Marrocchelli}}, \bibinfo {author}
  {\bibfnamefont {P.~A.}\ \bibnamefont {Madden}}, \bibinfo {author}
  {\bibfnamefont {P.}~\bibnamefont {Li}}, \ and\ \bibinfo {author}
  {\bibfnamefont {J.~T.~S.}\ \bibnamefont {Irvine}},\ }\href@noop {} {\bibfield
   {journal} {\bibinfo  {journal} {Chem Mater}\ }\textbf {\bibinfo {volume}
  {23}},\ \bibinfo {pages} {1356} (\bibinfo {year} {2011})}\BibitemShut
  {NoStop}%
\bibitem [{\citenamefont {Marrocchelli}\ \emph {et~al.}(2011)\citenamefont
  {Marrocchelli}, \citenamefont {Madden}, \citenamefont {Norberg},\ and\
  \citenamefont {Hull}}]{MarrocchelliEtAl_CM2011}%
  \BibitemOpen
  \bibfield  {author} {\bibinfo {author} {\bibfnamefont {D.}~\bibnamefont
  {Marrocchelli}}, \bibinfo {author} {\bibfnamefont {P.~A.}\ \bibnamefont
  {Madden}}, \bibinfo {author} {\bibfnamefont {S.~T.}\ \bibnamefont {Norberg}},
  \ and\ \bibinfo {author} {\bibfnamefont {S.}~\bibnamefont {Hull}},\
  }\href@noop {} {\bibfield  {journal} {\bibinfo  {journal} {Chem Mater}\
  }\textbf {\bibinfo {volume} {23}},\ \bibinfo {pages} {1365} (\bibinfo {year}
  {2011})}\BibitemShut {NoStop}%
\bibitem [{\citenamefont {Burbano}\ \emph {et~al.}(2012)\citenamefont
  {Burbano}, \citenamefont {Norberg}, \citenamefont {Hull}, \citenamefont
  {Eriksson}, \citenamefont {Marrocchelli}, \citenamefont {Madden},\ and\
  \citenamefont {Watson}}]{BurbanoEtAl_CM2012}%
  \BibitemOpen
  \bibfield  {author} {\bibinfo {author} {\bibfnamefont {M.}~\bibnamefont
  {Burbano}}, \bibinfo {author} {\bibfnamefont {S.~T.}\ \bibnamefont
  {Norberg}}, \bibinfo {author} {\bibfnamefont {S.}~\bibnamefont {Hull}},
  \bibinfo {author} {\bibfnamefont {S.~G.}\ \bibnamefont {Eriksson}}, \bibinfo
  {author} {\bibfnamefont {D.}~\bibnamefont {Marrocchelli}}, \bibinfo {author}
  {\bibfnamefont {P.~A.}\ \bibnamefont {Madden}}, \ and\ \bibinfo {author}
  {\bibfnamefont {G.~W.}\ \bibnamefont {Watson}},\ }\href@noop {} {\bibfield
  {journal} {\bibinfo  {journal} {Chem Mater}\ }\textbf {\bibinfo {volume}
  {24}},\ \bibinfo {pages} {222} (\bibinfo {year} {2012})}\BibitemShut
  {NoStop}%
\bibitem [{\citenamefont {Chen}\ \emph {et~al.}(2012)\citenamefont {Chen},
  \citenamefont {Sen},\ and\ \citenamefont {Kim}}]{ChenEtAl_CM2012}%
  \BibitemOpen
  \bibfield  {author} {\bibinfo {author} {\bibfnamefont {C.-T.}\ \bibnamefont
  {Chen}}, \bibinfo {author} {\bibfnamefont {S.}~\bibnamefont {Sen}}, \ and\
  \bibinfo {author} {\bibfnamefont {S.}~\bibnamefont {Kim}},\ }\href@noop {}
  {\bibfield  {journal} {\bibinfo  {journal} {Chem Mater}\ }\textbf {\bibinfo
  {volume} {24}},\ \bibinfo {pages} {3604} (\bibinfo {year}
  {2012})}\BibitemShut {NoStop}%
\bibitem [{\citenamefont {Marrocchelli}\ \emph {et~al.}(2013)\citenamefont
  {Marrocchelli}, \citenamefont {Bishop},\ and\ \citenamefont
  {Kilner}}]{MarrocchelliEtAl_JMCA2013}%
  \BibitemOpen
  \bibfield  {author} {\bibinfo {author} {\bibfnamefont {D.}~\bibnamefont
  {Marrocchelli}}, \bibinfo {author} {\bibfnamefont {S.~R.}\ \bibnamefont
  {Bishop}}, \ and\ \bibinfo {author} {\bibfnamefont {J.}~\bibnamefont
  {Kilner}},\ }\href@noop {} {\bibfield  {journal} {\bibinfo  {journal}
  {JOURNAL OF MATERIALS CHEMISTRY A}\ }\textbf {\bibinfo {volume} {1}},\
  \bibinfo {pages} {7673} (\bibinfo {year} {2013})}\BibitemShut {NoStop}%
\bibitem [{\citenamefont {Butler}\ \emph {et~al.}(1983)\citenamefont {Butler},
  \citenamefont {Catlow}, \citenamefont {Fender},\ and\ \citenamefont
  {Harding}}]{ButlerEtAl_SSI1983}%
  \BibitemOpen
  \bibfield  {author} {\bibinfo {author} {\bibfnamefont {V.}~\bibnamefont
  {Butler}}, \bibinfo {author} {\bibfnamefont {C.~R.~A.}\ \bibnamefont
  {Catlow}}, \bibinfo {author} {\bibfnamefont {B.~E.~F.}\ \bibnamefont
  {Fender}}, \ and\ \bibinfo {author} {\bibfnamefont {J.~H.}\ \bibnamefont
  {Harding}},\ }\href@noop {} {\bibfield  {journal} {\bibinfo  {journal} {Solid
  State Ionics}\ }\textbf {\bibinfo {volume} {8}},\ \bibinfo {pages} {109 }
  (\bibinfo {year} {1983})}\BibitemShut {NoStop}%
\bibitem [{\citenamefont {Kilner}(1983)}]{Kilner_SSI1983}%
  \BibitemOpen
  \bibfield  {author} {\bibinfo {author} {\bibfnamefont {J.}~\bibnamefont
  {Kilner}},\ }\href@noop {} {\bibfield  {journal} {\bibinfo  {journal} {Solid
  State Ionics}\ }\textbf {\bibinfo {volume} {8}},\ \bibinfo {pages} {201 }
  (\bibinfo {year} {1983})}\BibitemShut {NoStop}%
\bibitem [{\citenamefont {Balazs}\ and\ \citenamefont
  {Glass}(1995)}]{BalazsAndGlass_SSI1995}%
  \BibitemOpen
  \bibfield  {author} {\bibinfo {author} {\bibfnamefont {G.~B.}\ \bibnamefont
  {Balazs}}\ and\ \bibinfo {author} {\bibfnamefont {R.}~\bibnamefont {Glass}},\
  }\href@noop {} {\bibfield  {journal} {\bibinfo  {journal} {Solid State
  Ionics}\ }\textbf {\bibinfo {volume} {76}},\ \bibinfo {pages} {155} (\bibinfo
  {year} {1995})}\BibitemShut {NoStop}%
\bibitem [{\citenamefont {Hayashi}\ \emph {et~al.}(2000)\citenamefont
  {Hayashi}, \citenamefont {Sagawa}, \citenamefont {Inaba},\ and\ \citenamefont
  {Kawamura}}]{HayashiEtAL_SSI2000}%
  \BibitemOpen
  \bibfield  {author} {\bibinfo {author} {\bibfnamefont {H.}~\bibnamefont
  {Hayashi}}, \bibinfo {author} {\bibfnamefont {R.}~\bibnamefont {Sagawa}},
  \bibinfo {author} {\bibfnamefont {H.}~\bibnamefont {Inaba}}, \ and\ \bibinfo
  {author} {\bibfnamefont {K.}~\bibnamefont {Kawamura}},\ }\href@noop {}
  {\bibfield  {journal} {\bibinfo  {journal} {Solid State Ionics}\ }\textbf
  {\bibinfo {volume} {131}},\ \bibinfo {pages} {281 } (\bibinfo {year}
  {2000})}\BibitemShut {NoStop}%
\bibitem [{\citenamefont {Andersson}\ \emph {et~al.}(2006)\citenamefont
  {Andersson}, \citenamefont {Simak}, \citenamefont {Skorodumova},
  \citenamefont {Abrikosov},\ and\ \citenamefont
  {Johansson}}]{AnderssonEtAl_PNAS2006}%
  \BibitemOpen
  \bibfield  {author} {\bibinfo {author} {\bibfnamefont {D.~A.}\ \bibnamefont
  {Andersson}}, \bibinfo {author} {\bibfnamefont {S.~I.}\ \bibnamefont
  {Simak}}, \bibinfo {author} {\bibfnamefont {N.~V.}\ \bibnamefont
  {Skorodumova}}, \bibinfo {author} {\bibfnamefont {I.~A.}\ \bibnamefont
  {Abrikosov}}, \ and\ \bibinfo {author} {\bibfnamefont {B.}~\bibnamefont
  {Johansson}},\ }\href@noop {} {\bibfield  {journal} {\bibinfo  {journal}
  {Proc Natl Acad Sci U S A}\ }\textbf {\bibinfo {volume} {103}},\ \bibinfo
  {pages} {3518} (\bibinfo {year} {2006})}\BibitemShut {NoStop}%
\bibitem [{\citenamefont {Wang}\ \emph {et~al.}(2011)\citenamefont {Wang},
  \citenamefont {Lewis},\ and\ \citenamefont {Cormack}}]{WangEtAl_AC2011}%
  \BibitemOpen
  \bibfield  {author} {\bibinfo {author} {\bibfnamefont {B.}~\bibnamefont
  {Wang}}, \bibinfo {author} {\bibfnamefont {R.~J.}\ \bibnamefont {Lewis}}, \
  and\ \bibinfo {author} {\bibfnamefont {A.~N.}\ \bibnamefont {Cormack}},\
  }\href@noop {} {\bibfield  {journal} {\bibinfo  {journal} {Acta Mater}\
  }\textbf {\bibinfo {volume} {59}},\ \bibinfo {pages} {2035 } (\bibinfo {year}
  {2011})}\BibitemShut {NoStop}%
\bibitem [{\citenamefont {Rupp}(2012)}]{Rupp_SSI2012}%
  \BibitemOpen
  \bibfield  {author} {\bibinfo {author} {\bibfnamefont {J.~L.}\ \bibnamefont
  {Rupp}},\ }\href@noop {} {\bibfield  {journal} {\bibinfo  {journal} {Solid
  State Ionics}\ }\textbf {\bibinfo {volume} {207}},\ \bibinfo {pages} {1 }
  (\bibinfo {year} {2012})}\BibitemShut {NoStop}%
\bibitem [{\citenamefont {Burbano}\ \emph {et~al.}(2013)\citenamefont
  {Burbano}, \citenamefont {Marrocchelli},\ and\ \citenamefont
  {Watson}}]{BurbanoEtAl_JECR2013}%
  \BibitemOpen
  \bibfield  {author} {\bibinfo {author} {\bibfnamefont {M.}~\bibnamefont
  {Burbano}}, \bibinfo {author} {\bibfnamefont {D.}~\bibnamefont
  {Marrocchelli}}, \ and\ \bibinfo {author} {\bibfnamefont {G.}~\bibnamefont
  {Watson}},\ }\href@noop {} {\bibfield  {journal} {\bibinfo  {journal}
  {Journal of Electroceramics}\ ,\ \bibinfo {pages} {1}} (\bibinfo {year}
  {2013})}\BibitemShut {NoStop}%
\bibitem [{\citenamefont {Politova}\ and\ \citenamefont
  {Irvine}(2004)}]{PolitovaAndIrvine_SSI2004}%
  \BibitemOpen
  \bibfield  {author} {\bibinfo {author} {\bibfnamefont {T.}~\bibnamefont
  {Politova}}\ and\ \bibinfo {author} {\bibfnamefont {J.}~\bibnamefont
  {Irvine}},\ }\href@noop {} {\bibfield  {journal} {\bibinfo  {journal} {Solid
  State Ionics}\ }\textbf {\bibinfo {volume} {168}},\ \bibinfo {pages} {153}
  (\bibinfo {year} {2004})}\BibitemShut {NoStop}%
\bibitem [{\citenamefont {van Herle}\ \emph {et~al.}(1999)\citenamefont {van
  Herle}, \citenamefont {Seneviratne},\ and\ \citenamefont
  {McEvoy}}]{vanHerleEtAl_SECS1999}%
  \BibitemOpen
  \bibfield  {author} {\bibinfo {author} {\bibfnamefont {J.}~\bibnamefont {van
  Herle}}, \bibinfo {author} {\bibfnamefont {D.}~\bibnamefont {Seneviratne}}, \
  and\ \bibinfo {author} {\bibfnamefont {A.}~\bibnamefont {McEvoy}},\
  }\href@noop {} {\bibfield  {journal} {\bibinfo  {journal} {Journal of the
  European Ceramic Society}\ }\textbf {\bibinfo {volume} {19}},\ \bibinfo
  {pages} {837 } (\bibinfo {year} {1999})}\BibitemShut {NoStop}%
\bibitem [{\citenamefont {Singh}\ \emph {et~al.}(2013)\citenamefont {Singh},
  \citenamefont {Parkash},\ and\ \citenamefont {Kumar}}]{SinghEtAl_Ionics2013}%
  \BibitemOpen
  \bibfield  {author} {\bibinfo {author} {\bibfnamefont {N.}~\bibnamefont
  {Singh}}, \bibinfo {author} {\bibfnamefont {O.}~\bibnamefont {Parkash}}, \
  and\ \bibinfo {author} {\bibfnamefont {D.}~\bibnamefont {Kumar}},\
  }\href@noop {} {\bibfield  {journal} {\bibinfo  {journal} {Ionics}\ }\textbf
  {\bibinfo {volume} {19}},\ \bibinfo {pages} {165} (\bibinfo {year}
  {2013})}\BibitemShut {NoStop}%
\bibitem [{\citenamefont {Singh}\ \emph {et~al.}(2012)\citenamefont {Singh},
  \citenamefont {Singh}, \citenamefont {Kumar},\ and\ \citenamefont
  {Parkash}}]{SinghEtAl_Ionics2012}%
  \BibitemOpen
  \bibfield  {author} {\bibinfo {author} {\bibfnamefont {N.~K.}\ \bibnamefont
  {Singh}}, \bibinfo {author} {\bibfnamefont {P.}~\bibnamefont {Singh}},
  \bibinfo {author} {\bibfnamefont {D.}~\bibnamefont {Kumar}}, \ and\ \bibinfo
  {author} {\bibfnamefont {O.}~\bibnamefont {Parkash}},\ }\href@noop {}
  {\bibfield  {journal} {\bibinfo  {journal} {Ionics}\ }\textbf {\bibinfo
  {volume} {18}},\ \bibinfo {pages} {127} (\bibinfo {year} {2012})}\BibitemShut
  {NoStop}%
\bibitem [{\citenamefont {Omar}\ \emph {et~al.}(2009)\citenamefont {Omar},
  \citenamefont {Wachsman}, \citenamefont {Jones},\ and\ \citenamefont
  {Nino}}]{OmarEtAl_JACS2009}%
  \BibitemOpen
  \bibfield  {author} {\bibinfo {author} {\bibfnamefont {S.}~\bibnamefont
  {Omar}}, \bibinfo {author} {\bibfnamefont {E.~D.}\ \bibnamefont {Wachsman}},
  \bibinfo {author} {\bibfnamefont {J.~L.}\ \bibnamefont {Jones}}, \ and\
  \bibinfo {author} {\bibfnamefont {J.~C.}\ \bibnamefont {Nino}},\ }\href@noop
  {} {\bibfield  {journal} {\bibinfo  {journal} {J Am Ceram Soc}\ }\textbf
  {\bibinfo {volume} {92}},\ \bibinfo {pages} {2674} (\bibinfo {year}
  {2009})}\BibitemShut {NoStop}%
\bibitem [{\citenamefont {Omar}\ \emph {et~al.}(2007)\citenamefont {Omar},
  \citenamefont {Wachsman},\ and\ \citenamefont {Nino}}]{OmarEtAl_APL2007}%
  \BibitemOpen
  \bibfield  {author} {\bibinfo {author} {\bibfnamefont {S.}~\bibnamefont
  {Omar}}, \bibinfo {author} {\bibfnamefont {E.~D.}\ \bibnamefont {Wachsman}},
  \ and\ \bibinfo {author} {\bibfnamefont {J.~C.}\ \bibnamefont {Nino}},\
  }\href@noop {} {\bibfield  {journal} {\bibinfo  {journal} {Appl Phys Lett}\
  }\textbf {\bibinfo {volume} {91}},\ \bibinfo {pages} {144106} (\bibinfo
  {year} {2007})}\BibitemShut {NoStop}%
\bibitem [{\citenamefont {Sha}\ \emph {et~al.}(2006)\citenamefont {Sha},
  \citenamefont {L\"{u}}, \citenamefont {Huang}, \citenamefont {Miao},
  \citenamefont {Jia}, \citenamefont {Xin},\ and\ \citenamefont
  {Su}}]{ShaEtAl_JAC2006}%
  \BibitemOpen
  \bibfield  {author} {\bibinfo {author} {\bibfnamefont {X.}~\bibnamefont
  {Sha}}, \bibinfo {author} {\bibfnamefont {Z.}~\bibnamefont {L\"{u}}},
  \bibinfo {author} {\bibfnamefont {X.}~\bibnamefont {Huang}}, \bibinfo
  {author} {\bibfnamefont {J.}~\bibnamefont {Miao}}, \bibinfo {author}
  {\bibfnamefont {L.}~\bibnamefont {Jia}}, \bibinfo {author} {\bibfnamefont
  {X.}~\bibnamefont {Xin}}, \ and\ \bibinfo {author} {\bibfnamefont
  {W.}~\bibnamefont {Su}},\ }\href@noop {} {\bibfield  {journal} {\bibinfo
  {journal} {Journal of Alloys and Compounds}\ }\textbf {\bibinfo {volume}
  {424}},\ \bibinfo {pages} {315 } (\bibinfo {year} {2006})}\BibitemShut
  {NoStop}%
\bibitem [{\citenamefont {Sha}\ \emph {et~al.}(2007)\citenamefont {Sha},
  \citenamefont {L\"{u}}, \citenamefont {Huang}, \citenamefont {Miao},
  \citenamefont {Ding}, \citenamefont {Xin},\ and\ \citenamefont
  {Su}}]{ShaEtAl_JAC2007}%
  \BibitemOpen
  \bibfield  {author} {\bibinfo {author} {\bibfnamefont {X.}~\bibnamefont
  {Sha}}, \bibinfo {author} {\bibfnamefont {Z.}~\bibnamefont {L\"{u}}},
  \bibinfo {author} {\bibfnamefont {X.}~\bibnamefont {Huang}}, \bibinfo
  {author} {\bibfnamefont {J.}~\bibnamefont {Miao}}, \bibinfo {author}
  {\bibfnamefont {Z.}~\bibnamefont {Ding}}, \bibinfo {author} {\bibfnamefont
  {X.}~\bibnamefont {Xin}}, \ and\ \bibinfo {author} {\bibfnamefont
  {W.}~\bibnamefont {Su}},\ }\href@noop {} {\bibfield  {journal} {\bibinfo
  {journal} {J Alloys Compd}\ }\textbf {\bibinfo {volume} {428}},\ \bibinfo
  {pages} {59 } (\bibinfo {year} {2007})}\BibitemShut {NoStop}%
\bibitem [{\citenamefont {Dikmen}\ \emph {et~al.}(2010)\citenamefont {Dikmen},
  \citenamefont {Aslanbay}, \citenamefont {Dikmen},\ and\ \citenamefont
  {Sahin}}]{DikmenEtAl_JPS2010}%
  \BibitemOpen
  \bibfield  {author} {\bibinfo {author} {\bibfnamefont {S.}~\bibnamefont
  {Dikmen}}, \bibinfo {author} {\bibfnamefont {H.}~\bibnamefont {Aslanbay}},
  \bibinfo {author} {\bibfnamefont {E.}~\bibnamefont {Dikmen}}, \ and\ \bibinfo
  {author} {\bibfnamefont {O.}~\bibnamefont {Sahin}},\ }\href@noop {}
  {\bibfield  {journal} {\bibinfo  {journal} {J Power Sources}\ }\textbf
  {\bibinfo {volume} {195}},\ \bibinfo {pages} {2488 } (\bibinfo {year}
  {2010})}\BibitemShut {NoStop}%
\bibitem [{\citenamefont {Guan}\ \emph {et~al.}(2008)\citenamefont {Guan},
  \citenamefont {Zhou}, \citenamefont {Liu}, \citenamefont {Wang},\ and\
  \citenamefont {Zhang}}]{GuanEtAl_MRB2008}%
  \BibitemOpen
  \bibfield  {author} {\bibinfo {author} {\bibfnamefont {X.}~\bibnamefont
  {Guan}}, \bibinfo {author} {\bibfnamefont {H.}~\bibnamefont {Zhou}}, \bibinfo
  {author} {\bibfnamefont {Z.}~\bibnamefont {Liu}}, \bibinfo {author}
  {\bibfnamefont {Y.}~\bibnamefont {Wang}}, \ and\ \bibinfo {author}
  {\bibfnamefont {J.}~\bibnamefont {Zhang}},\ }\href@noop {} {\bibfield
  {journal} {\bibinfo  {journal} {Materials Research Bulletin}\ }\textbf
  {\bibinfo {volume} {43}},\ \bibinfo {pages} {1046 } (\bibinfo {year}
  {2008})}\BibitemShut {NoStop}%
\bibitem [{\citenamefont {Ayawanna}\ \emph {et~al.}(2009)\citenamefont
  {Ayawanna}, \citenamefont {Wattanasiriwech}, \citenamefont
  {Wattanasiriwech},\ and\ \citenamefont
  {Aungkavattana}}]{AyawannaEtAl_SSI2009}%
  \BibitemOpen
  \bibfield  {author} {\bibinfo {author} {\bibfnamefont {J.}~\bibnamefont
  {Ayawanna}}, \bibinfo {author} {\bibfnamefont {D.}~\bibnamefont
  {Wattanasiriwech}}, \bibinfo {author} {\bibfnamefont {S.}~\bibnamefont
  {Wattanasiriwech}}, \ and\ \bibinfo {author} {\bibfnamefont {P.}~\bibnamefont
  {Aungkavattana}},\ }\href@noop {} {\bibfield  {journal} {\bibinfo  {journal}
  {Solid State Ionics}\ }\textbf {\bibinfo {volume} {180}},\ \bibinfo {pages}
  {1388 } (\bibinfo {year} {2009})}\BibitemShut {NoStop}%
\bibitem [{\citenamefont {Dholabhai}\ \emph {et~al.}(2011)\citenamefont
  {Dholabhai}, \citenamefont {Adams}, \citenamefont {Crozier},\ and\
  \citenamefont {Sharma}}]{DholabhaiEtAl_JMC2011}%
  \BibitemOpen
  \bibfield  {author} {\bibinfo {author} {\bibfnamefont {P.~P.}\ \bibnamefont
  {Dholabhai}}, \bibinfo {author} {\bibfnamefont {J.~B.}\ \bibnamefont
  {Adams}}, \bibinfo {author} {\bibfnamefont {P.~A.}\ \bibnamefont {Crozier}},
  \ and\ \bibinfo {author} {\bibfnamefont {R.}~\bibnamefont {Sharma}},\
  }\href@noop {} {\bibfield  {journal} {\bibinfo  {journal} {J Mater Chem}\
  }\textbf {\bibinfo {volume} {21}},\ \bibinfo {pages} {18991} (\bibinfo {year}
  {2011})}\BibitemShut {NoStop}%
\bibitem [{\citenamefont {Ralph}\ \emph {et~al.}(1997)\citenamefont {Ralph},
  \citenamefont {Przydatek}, \citenamefont {Kilner},\ and\ \citenamefont
  {Seguelong}}]{RalphEtAl_BBG1997}%
  \BibitemOpen
  \bibfield  {author} {\bibinfo {author} {\bibfnamefont {J.}~\bibnamefont
  {Ralph}}, \bibinfo {author} {\bibfnamefont {J.}~\bibnamefont {Przydatek}},
  \bibinfo {author} {\bibfnamefont {J.}~\bibnamefont {Kilner}}, \ and\ \bibinfo
  {author} {\bibfnamefont {T.}~\bibnamefont {Seguelong}},\ }\href@noop {}
  {\bibfield  {journal} {\bibinfo  {journal} {Berichte der
  Bunsen-Gesellschaft}\ }\textbf {\bibinfo {volume} {101}},\ \bibinfo {pages}
  {1403} (\bibinfo {year} {1997})}\BibitemShut {NoStop}%
\bibitem [{\citenamefont {Kasse}\ and\ \citenamefont
  {Nino}(2013)}]{KasseAndNino_JAC2013}%
  \BibitemOpen
  \bibfield  {author} {\bibinfo {author} {\bibfnamefont {R.~M.}\ \bibnamefont
  {Kasse}}\ and\ \bibinfo {author} {\bibfnamefont {J.~C.}\ \bibnamefont
  {Nino}},\ }\href@noop {} {\bibfield  {journal} {\bibinfo  {journal} {Journal
  of Alloys and Compounds}\ }\textbf {\bibinfo {volume} {575}},\ \bibinfo
  {pages} {399 } (\bibinfo {year} {2013})}\BibitemShut {NoStop}%
\bibitem [{\citenamefont {Kim}(1989)}]{Kim_JACS1989}%
  \BibitemOpen
  \bibfield  {author} {\bibinfo {author} {\bibfnamefont {D.-J.}\ \bibnamefont
  {Kim}},\ }\href@noop {} {\bibfield  {journal} {\bibinfo  {journal} {J Am
  Ceram Soc}\ }\textbf {\bibinfo {volume} {72}},\ \bibinfo {pages} {1415}
  (\bibinfo {year} {1989})}\BibitemShut {NoStop}%
\bibitem [{Note2()}]{Note2}%
  \BibitemOpen
  \bibinfo {note} {We note here that the radii of these three elements (1.079,
  1.093, 1.109 \r A\ for Sm, Pm and Nd, respectively \cite {Shannon_ACA1976})
  are all very similar and almost within the associated experimental
  error.}\BibitemShut {Stop}%
\bibitem [{\citenamefont {Omar}\ \emph {et~al.}(2006)\citenamefont {Omar},
  \citenamefont {Wachsman},\ and\ \citenamefont {Nino}}]{OmarEtAl_SSI2006}%
  \BibitemOpen
  \bibfield  {author} {\bibinfo {author} {\bibfnamefont {S.}~\bibnamefont
  {Omar}}, \bibinfo {author} {\bibfnamefont {E.~D.}\ \bibnamefont {Wachsman}},
  \ and\ \bibinfo {author} {\bibfnamefont {J.~C.}\ \bibnamefont {Nino}},\
  }\href@noop {} {\bibfield  {journal} {\bibinfo  {journal} {Solid State
  Ionics}\ }\textbf {\bibinfo {volume} {177}},\ \bibinfo {pages} {3199 }
  (\bibinfo {year} {2006})}\BibitemShut {NoStop}%
\bibitem [{\citenamefont {Chroneos}\ \emph {et~al.}(2011)\citenamefont
  {Chroneos}, \citenamefont {Yildiz}, \citenamefont {Tarancon}, \citenamefont
  {Parfitt},\ and\ \citenamefont {Kilner}}]{ChroneosEtAl_EES2011}%
  \BibitemOpen
  \bibfield  {author} {\bibinfo {author} {\bibfnamefont {A.}~\bibnamefont
  {Chroneos}}, \bibinfo {author} {\bibfnamefont {B.}~\bibnamefont {Yildiz}},
  \bibinfo {author} {\bibfnamefont {A.}~\bibnamefont {Tarancon}}, \bibinfo
  {author} {\bibfnamefont {D.}~\bibnamefont {Parfitt}}, \ and\ \bibinfo
  {author} {\bibfnamefont {J.~A.}\ \bibnamefont {Kilner}},\ }\href@noop {}
  {\bibfield  {journal} {\bibinfo  {journal} {Energy Environ. Sci.}\ }\textbf
  {\bibinfo {volume} {4}},\ \bibinfo {pages} {2774} (\bibinfo {year}
  {2011})}\BibitemShut {NoStop}%
\bibitem [{\citenamefont {Balducci}\ \emph {et~al.}(2000)\citenamefont
  {Balducci}, \citenamefont {Islam}, \citenamefont {Ka$\check{s}$par},
  \citenamefont {Fornasiero},\ and\ \citenamefont
  {Graziani}}]{BalducciEtAl_CM2000}%
  \BibitemOpen
  \bibfield  {author} {\bibinfo {author} {\bibfnamefont {G.}~\bibnamefont
  {Balducci}}, \bibinfo {author} {\bibfnamefont {M.~S.}\ \bibnamefont {Islam}},
  \bibinfo {author} {\bibfnamefont {J.}~\bibnamefont {Ka$\check{s}$par}},
  \bibinfo {author} {\bibfnamefont {P.}~\bibnamefont {Fornasiero}}, \ and\
  \bibinfo {author} {\bibfnamefont {M.}~\bibnamefont {Graziani}},\ }\href@noop
  {} {\bibfield  {journal} {\bibinfo  {journal} {Chem Mater}\ }\textbf
  {\bibinfo {volume} {12}},\ \bibinfo {pages} {677} (\bibinfo {year}
  {2000})}\BibitemShut {NoStop}%
\bibitem [{\citenamefont {Goodenough}(2003)}]{Goodenough_ARMR2003}%
  \BibitemOpen
  \bibfield  {author} {\bibinfo {author} {\bibfnamefont {J.~B.}\ \bibnamefont
  {Goodenough}},\ }\href@noop {} {\bibfield  {journal} {\bibinfo  {journal}
  {Annu Rev Mater Res}\ }\textbf {\bibinfo {volume} {33}},\ \bibinfo {pages}
  {91} (\bibinfo {year} {2003})}\BibitemShut {NoStop}%
\bibitem [{\citenamefont {Gobel}\ \emph {et~al.}(2010)\citenamefont {Gobel},
  \citenamefont {Gregori}, \citenamefont {Guo},\ and\ \citenamefont
  {Maier}}]{GobelEtAl_PCCP2010}%
  \BibitemOpen
  \bibfield  {author} {\bibinfo {author} {\bibfnamefont {M.~C.}\ \bibnamefont
  {Gobel}}, \bibinfo {author} {\bibfnamefont {G.}~\bibnamefont {Gregori}},
  \bibinfo {author} {\bibfnamefont {X.}~\bibnamefont {Guo}}, \ and\ \bibinfo
  {author} {\bibfnamefont {J.}~\bibnamefont {Maier}},\ }\href@noop {}
  {\bibfield  {journal} {\bibinfo  {journal} {Phys Chem Chem Phys}\ }\textbf
  {\bibinfo {volume} {12}},\ \bibinfo {pages} {14351} (\bibinfo {year}
  {2010})}\BibitemShut {NoStop}%
\bibitem [{\citenamefont {Da~Silva}\ \emph {et~al.}(2007)\citenamefont
  {Da~Silva}, \citenamefont {Ganduglia-Pirovano}, \citenamefont {Sauer},
  \citenamefont {Bayer},\ and\ \citenamefont {Kresse}}]{DaSilvaEtAl_PRB2007}%
  \BibitemOpen
  \bibfield  {author} {\bibinfo {author} {\bibfnamefont {J.~L.~F.}\
  \bibnamefont {Da~Silva}}, \bibinfo {author} {\bibfnamefont {M.~V.}\
  \bibnamefont {Ganduglia-Pirovano}}, \bibinfo {author} {\bibfnamefont
  {J.}~\bibnamefont {Sauer}}, \bibinfo {author} {\bibfnamefont
  {V.}~\bibnamefont {Bayer}}, \ and\ \bibinfo {author} {\bibfnamefont
  {G.}~\bibnamefont {Kresse}},\ }\href@noop {} {\bibfield  {journal} {\bibinfo
  {journal} {Physical Review B}\ }\textbf {\bibinfo {volume} {75}},\ \bibinfo
  {pages} {045121} (\bibinfo {year} {2007})}\BibitemShut {NoStop}%
\bibitem [{\citenamefont {Gillen}\ \emph {et~al.}(2013)\citenamefont {Gillen},
  \citenamefont {Clark},\ and\ \citenamefont {Robertson}}]{GillenEtAl_PRB2013}%
  \BibitemOpen
  \bibfield  {author} {\bibinfo {author} {\bibfnamefont {R.}~\bibnamefont
  {Gillen}}, \bibinfo {author} {\bibfnamefont {S.~J.}\ \bibnamefont {Clark}}, \
  and\ \bibinfo {author} {\bibfnamefont {J.}~\bibnamefont {Robertson}},\
  }\href@noop {} {\bibfield  {journal} {\bibinfo  {journal} {Physical Review
  B}\ }\textbf {\bibinfo {volume} {87}},\ \bibinfo {pages} {125116} (\bibinfo
  {year} {2013})}\BibitemShut {NoStop}%
\bibitem [{\citenamefont {Ganduglia-Pirovano}\ \emph
  {et~al.}(2007)\citenamefont {Ganduglia-Pirovano}, \citenamefont {Hofmann},\
  and\ \citenamefont {Sauer}}]{Ganduglia-PirovanoEtAl_SSR2007}%
  \BibitemOpen
  \bibfield  {author} {\bibinfo {author} {\bibfnamefont {M.~V.}\ \bibnamefont
  {Ganduglia-Pirovano}}, \bibinfo {author} {\bibfnamefont {A.}~\bibnamefont
  {Hofmann}}, \ and\ \bibinfo {author} {\bibfnamefont {J.}~\bibnamefont
  {Sauer}},\ }\href@noop {} {\bibfield  {journal} {\bibinfo  {journal} {Surface
  Science Reports}\ }\textbf {\bibinfo {volume} {62}},\ \bibinfo {pages} {219 }
  (\bibinfo {year} {2007})}\BibitemShut {NoStop}%
\bibitem [{\citenamefont {Keating}\ \emph {et~al.}(2009)\citenamefont
  {Keating}, \citenamefont {Scanlon},\ and\ \citenamefont
  {Watson}}]{KeatingEtAl_JPCM2009}%
  \BibitemOpen
  \bibfield  {author} {\bibinfo {author} {\bibfnamefont {P.~R.}\ \bibnamefont
  {Keating}}, \bibinfo {author} {\bibfnamefont {D.~O.}\ \bibnamefont
  {Scanlon}}, \ and\ \bibinfo {author} {\bibfnamefont {G.~W.}\ \bibnamefont
  {Watson}},\ }\href@noop {} {\bibfield  {journal} {\bibinfo  {journal} {J Phys
  : Condens Matter}\ }\textbf {\bibinfo {volume} {21}},\ \bibinfo {pages}
  {405502} (\bibinfo {year} {2009})}\BibitemShut {NoStop}%
\bibitem [{\citenamefont {Keating}\ \emph {et~al.}(2012)\citenamefont
  {Keating}, \citenamefont {Scanlon}, \citenamefont {Morgan}, \citenamefont
  {Galea},\ and\ \citenamefont {Watson}}]{KeatingEtAl_JPC2012}%
  \BibitemOpen
  \bibfield  {author} {\bibinfo {author} {\bibfnamefont {P.~R.~L.}\
  \bibnamefont {Keating}}, \bibinfo {author} {\bibfnamefont {D.~O.}\
  \bibnamefont {Scanlon}}, \bibinfo {author} {\bibfnamefont {B.~J.}\
  \bibnamefont {Morgan}}, \bibinfo {author} {\bibfnamefont {N.~M.}\
  \bibnamefont {Galea}}, \ and\ \bibinfo {author} {\bibfnamefont {G.~W.}\
  \bibnamefont {Watson}},\ }\href@noop {} {\bibfield  {journal} {\bibinfo
  {journal} {The Journal of Physical Chemistry C}\ }\textbf {\bibinfo {volume}
  {116}},\ \bibinfo {pages} {2443} (\bibinfo {year} {2012})}\BibitemShut
  {NoStop}%
\bibitem [{\citenamefont {Keating}\ \emph {et~al.}(2013)\citenamefont
  {Keating}, \citenamefont {Scanlon},\ and\ \citenamefont
  {Watson}}]{KeatingEtAl_JMCC2013}%
  \BibitemOpen
  \bibfield  {author} {\bibinfo {author} {\bibfnamefont {P.~R.~L.}\
  \bibnamefont {Keating}}, \bibinfo {author} {\bibfnamefont {D.~O.}\
  \bibnamefont {Scanlon}}, \ and\ \bibinfo {author} {\bibfnamefont {G.~W.}\
  \bibnamefont {Watson}},\ }\href@noop {} {\bibfield  {journal} {\bibinfo
  {journal} {J. Mater. Chem. C}\ }\textbf {\bibinfo {volume} {1}},\ \bibinfo
  {pages} {1093} (\bibinfo {year} {2013})}\BibitemShut {NoStop}%
\bibitem [{\citenamefont {Nolan}\ \emph {et~al.}(2006)\citenamefont {Nolan},
  \citenamefont {Fearon},\ and\ \citenamefont {Watson}}]{NolanEtAL_SSI2006}%
  \BibitemOpen
  \bibfield  {author} {\bibinfo {author} {\bibfnamefont {M.}~\bibnamefont
  {Nolan}}, \bibinfo {author} {\bibfnamefont {J.~E.}\ \bibnamefont {Fearon}}, \
  and\ \bibinfo {author} {\bibfnamefont {G.~W.}\ \bibnamefont {Watson}},\
  }\href@noop {} {\bibfield  {journal} {\bibinfo  {journal} {Solid State
  Ionics}\ }\textbf {\bibinfo {volume} {177}},\ \bibinfo {pages} {3069 }
  (\bibinfo {year} {2006})}\BibitemShut {NoStop}%
\bibitem [{\citenamefont {Andersson}\ \emph {et~al.}(2007)\citenamefont
  {Andersson}, \citenamefont {Simak}, \citenamefont {Johansson}, \citenamefont
  {Abrikosov},\ and\ \citenamefont {Skorodumova}}]{AnderssonEtAl_PhysRevB2007}%
  \BibitemOpen
  \bibfield  {author} {\bibinfo {author} {\bibfnamefont {D.~A.}\ \bibnamefont
  {Andersson}}, \bibinfo {author} {\bibfnamefont {S.~I.}\ \bibnamefont
  {Simak}}, \bibinfo {author} {\bibfnamefont {B.}~\bibnamefont {Johansson}},
  \bibinfo {author} {\bibfnamefont {I.~A.}\ \bibnamefont {Abrikosov}}, \ and\
  \bibinfo {author} {\bibfnamefont {N.~V.}\ \bibnamefont {Skorodumova}},\
  }\href@noop {} {\bibfield  {journal} {\bibinfo  {journal} {Physical Review
  B}\ }\textbf {\bibinfo {volume} {75}},\ \bibinfo {pages} {035109} (\bibinfo
  {year} {2007})}\BibitemShut {NoStop}%
\bibitem [{\citenamefont {Madden}\ and\ \citenamefont
  {Wilson}(1996)}]{MaddenAndWilson_CSR1996}%
  \BibitemOpen
  \bibfield  {author} {\bibinfo {author} {\bibfnamefont {P.~A.}\ \bibnamefont
  {Madden}}\ and\ \bibinfo {author} {\bibfnamefont {M.}~\bibnamefont
  {Wilson}},\ }\href@noop {} {\bibfield  {journal} {\bibinfo  {journal} {Chem
  Soc Rev}\ }\textbf {\bibinfo {volume} {25}},\ \bibinfo {pages} {339}
  (\bibinfo {year} {1996})}\BibitemShut {NoStop}%
\bibitem [{\citenamefont {Marrocchelli}\ \emph {et~al.}(2012)\citenamefont
  {Marrocchelli}, \citenamefont {Bishop}, \citenamefont {Tuller},\ and\
  \citenamefont {Yildiz}}]{MarrocchelliEtAl_AFM2012}%
  \BibitemOpen
  \bibfield  {author} {\bibinfo {author} {\bibfnamefont {D.}~\bibnamefont
  {Marrocchelli}}, \bibinfo {author} {\bibfnamefont {S.~R.}\ \bibnamefont
  {Bishop}}, \bibinfo {author} {\bibfnamefont {H.~L.}\ \bibnamefont {Tuller}},
  \ and\ \bibinfo {author} {\bibfnamefont {B.}~\bibnamefont {Yildiz}},\
  }\href@noop {} {\bibfield  {journal} {\bibinfo  {journal} {Adv Funct Mater}\
  }\textbf {\bibinfo {volume} {22}},\ \bibinfo {pages} {1958} (\bibinfo {year}
  {2012})}\BibitemShut {NoStop}%
\bibitem [{\citenamefont {Marrocchelli}\ \emph {et~al.}(2010)\citenamefont
  {Marrocchelli}, \citenamefont {Salanne},\ and\ \citenamefont
  {Madden}}]{Marrocchelli_JPCM2010}%
  \BibitemOpen
  \bibfield  {author} {\bibinfo {author} {\bibfnamefont {D.}~\bibnamefont
  {Marrocchelli}}, \bibinfo {author} {\bibfnamefont {M.}~\bibnamefont
  {Salanne}}, \ and\ \bibinfo {author} {\bibfnamefont {P.~A.}\ \bibnamefont
  {Madden}},\ }\href@noop {} {\bibfield  {journal} {\bibinfo  {journal} {J Phys
  : Condens Matter}\ }\textbf {\bibinfo {volume} {22}},\ \bibinfo {pages}
  {152102} (\bibinfo {year} {2010})}\BibitemShut {NoStop}%
\bibitem [{\citenamefont {Marrocchelli}\ \emph
  {et~al.}(2009{\natexlab{a}})\citenamefont {Marrocchelli}, \citenamefont
  {Salanne}, \citenamefont {Madden}, \citenamefont {Simon},\ and\ \citenamefont
  {Turq}}]{MarrocchelliEtAl_MP2009}%
  \BibitemOpen
  \bibfield  {author} {\bibinfo {author} {\bibfnamefont {D.}~\bibnamefont
  {Marrocchelli}}, \bibinfo {author} {\bibfnamefont {M.}~\bibnamefont
  {Salanne}}, \bibinfo {author} {\bibfnamefont {P.~A.}\ \bibnamefont {Madden}},
  \bibinfo {author} {\bibfnamefont {C.}~\bibnamefont {Simon}}, \ and\ \bibinfo
  {author} {\bibfnamefont {P.}~\bibnamefont {Turq}},\ }\href@noop {} {\bibfield
   {journal} {\bibinfo  {journal} {Mol Phys}\ }\textbf {\bibinfo {volume}
  {107}},\ \bibinfo {pages} {443 } (\bibinfo {year}
  {2009}{\natexlab{a}})}\BibitemShut {NoStop}%
\bibitem [{\citenamefont {Marrocchelli}\ \emph
  {et~al.}(2009{\natexlab{b}})\citenamefont {Marrocchelli}, \citenamefont
  {Madden}, \citenamefont {Norberg},\ and\ \citenamefont
  {Hull}}]{MarrocchelliEtAl_JPCM2009}%
  \BibitemOpen
  \bibfield  {author} {\bibinfo {author} {\bibfnamefont {D.}~\bibnamefont
  {Marrocchelli}}, \bibinfo {author} {\bibfnamefont {P.~A.}\ \bibnamefont
  {Madden}}, \bibinfo {author} {\bibfnamefont {S.~T.}\ \bibnamefont {Norberg}},
  \ and\ \bibinfo {author} {\bibfnamefont {S.}~\bibnamefont {Hull}},\
  }\href@noop {} {\bibfield  {journal} {\bibinfo  {journal} {J Phys : Condens
  Matter}\ }\textbf {\bibinfo {volume} {21}},\ \bibinfo {pages} {405403}
  (\bibinfo {year} {2009}{\natexlab{b}})}\BibitemShut {NoStop}%
\bibitem [{\citenamefont {Wilson}\ \emph {et~al.}(2004)\citenamefont {Wilson},
  \citenamefont {Jahn},\ and\ \citenamefont {Madden}}]{WilsonEtAl_JPCM2004}%
  \BibitemOpen
  \bibfield  {author} {\bibinfo {author} {\bibfnamefont {M.}~\bibnamefont
  {Wilson}}, \bibinfo {author} {\bibfnamefont {S.}~\bibnamefont {Jahn}}, \ and\
  \bibinfo {author} {\bibfnamefont {P.~A.}\ \bibnamefont {Madden}},\
  }\href@noop {} {\bibfield  {journal} {\bibinfo  {journal} {J Phys : Condens
  Matter}\ }\textbf {\bibinfo {volume} {16}},\ \bibinfo {pages} {S2795}
  (\bibinfo {year} {2004})}\BibitemShut {NoStop}%
\bibitem [{\citenamefont {Burbano}\ \emph {et~al.}(2011)\citenamefont
  {Burbano}, \citenamefont {Marrocchelli}, \citenamefont {Yildiz},
  \citenamefont {Tuller}, \citenamefont {Norberg}, \citenamefont {Hull},
  \citenamefont {Madden},\ and\ \citenamefont {Watson}}]{BurbanoEtAl_JPCM2011}%
  \BibitemOpen
  \bibfield  {author} {\bibinfo {author} {\bibfnamefont {M.}~\bibnamefont
  {Burbano}}, \bibinfo {author} {\bibfnamefont {D.}~\bibnamefont
  {Marrocchelli}}, \bibinfo {author} {\bibfnamefont {B.}~\bibnamefont
  {Yildiz}}, \bibinfo {author} {\bibfnamefont {H.~L.}\ \bibnamefont {Tuller}},
  \bibinfo {author} {\bibfnamefont {S.~T.}\ \bibnamefont {Norberg}}, \bibinfo
  {author} {\bibfnamefont {S.}~\bibnamefont {Hull}}, \bibinfo {author}
  {\bibfnamefont {P.~A.}\ \bibnamefont {Madden}}, \ and\ \bibinfo {author}
  {\bibfnamefont {G.~W.}\ \bibnamefont {Watson}},\ }\href@noop {} {\bibfield
  {journal} {\bibinfo  {journal} {J Phys : Condens Matter}\ }\textbf {\bibinfo
  {volume} {23}},\ \bibinfo {pages} {255402} (\bibinfo {year}
  {2011})}\BibitemShut {NoStop}%
\bibitem [{\citenamefont {Heaton}\ \emph
  {et~al.}(2006{\natexlab{a}})\citenamefont {Heaton}, \citenamefont {Brookes},
  \citenamefont {Madden}, \citenamefont {Salanne}, \citenamefont {Simon},\ and\
  \citenamefont {Turq}}]{HeatonEtAl_JPCB2006}%
  \BibitemOpen
  \bibfield  {author} {\bibinfo {author} {\bibfnamefont {R.}~\bibnamefont
  {Heaton}}, \bibinfo {author} {\bibfnamefont {R.}~\bibnamefont {Brookes}},
  \bibinfo {author} {\bibfnamefont {P.}~\bibnamefont {Madden}}, \bibinfo
  {author} {\bibfnamefont {M.}~\bibnamefont {Salanne}}, \bibinfo {author}
  {\bibfnamefont {C.}~\bibnamefont {Simon}}, \ and\ \bibinfo {author}
  {\bibfnamefont {P.}~\bibnamefont {Turq}},\ }\href@noop {} {\bibfield
  {journal} {\bibinfo  {journal} {J Phys Chem B}\ }\textbf {\bibinfo {volume}
  {110}},\ \bibinfo {pages} {11454} (\bibinfo {year}
  {2006}{\natexlab{a}})}\BibitemShut {NoStop}%
\bibitem [{\citenamefont {Salanne}\ \emph {et~al.}(2009)\citenamefont
  {Salanne}, \citenamefont {Simon}, \citenamefont {Turq},\ and\ \citenamefont
  {Madden}}]{SalanneEtAl_JFC2009}%
  \BibitemOpen
  \bibfield  {author} {\bibinfo {author} {\bibfnamefont {M.}~\bibnamefont
  {Salanne}}, \bibinfo {author} {\bibfnamefont {C.}~\bibnamefont {Simon}},
  \bibinfo {author} {\bibfnamefont {P.}~\bibnamefont {Turq}}, \ and\ \bibinfo
  {author} {\bibfnamefont {P.}~\bibnamefont {Madden}},\ }\href@noop {}
  {\bibfield  {journal} {\bibinfo  {journal} {J Fluorine Chem}\ }\textbf
  {\bibinfo {volume} {130}},\ \bibinfo {pages} {38} (\bibinfo {year}
  {2009})}\BibitemShut {NoStop}%
\bibitem [{\citenamefont {Salanne}\ \emph {et~al.}(2012)\citenamefont
  {Salanne}, \citenamefont {Marrocchelli},\ and\ \citenamefont
  {Watson}}]{SalanneEtAl_JPCC2012}%
  \BibitemOpen
  \bibfield  {author} {\bibinfo {author} {\bibfnamefont {M.}~\bibnamefont
  {Salanne}}, \bibinfo {author} {\bibfnamefont {D.}~\bibnamefont
  {Marrocchelli}}, \ and\ \bibinfo {author} {\bibfnamefont {G.~W.}\
  \bibnamefont {Watson}},\ }\href@noop {} {\bibfield  {journal} {\bibinfo
  {journal} {The Journal of Physical Chemistry C}\ }\textbf {\bibinfo {volume}
  {116}},\ \bibinfo {pages} {18618} (\bibinfo {year} {2012})}\BibitemShut
  {NoStop}%
\bibitem [{\citenamefont {Martyna}\ \emph {et~al.}(1994)\citenamefont
  {Martyna}, \citenamefont {Tobias},\ and\ \citenamefont
  {Klein}}]{MartynaEtAl_JCP1994}%
  \BibitemOpen
  \bibfield  {author} {\bibinfo {author} {\bibfnamefont {G.~J.}\ \bibnamefont
  {Martyna}}, \bibinfo {author} {\bibfnamefont {D.~J.}\ \bibnamefont {Tobias}},
  \ and\ \bibinfo {author} {\bibfnamefont {M.~L.}\ \bibnamefont {Klein}},\
  }\href@noop {} {\bibfield  {journal} {\bibinfo  {journal} {J Chem Phys}\
  }\textbf {\bibinfo {volume} {101}},\ \bibinfo {pages} {4177} (\bibinfo {year}
  {1994})}\BibitemShut {NoStop}%
\bibitem [{\citenamefont {Ewald}(1921)}]{Ewald_AP1921}%
  \BibitemOpen
  \bibfield  {author} {\bibinfo {author} {\bibfnamefont {P.~P.}\ \bibnamefont
  {Ewald}},\ }\href@noop {} {\bibfield  {journal} {\bibinfo  {journal} {Annalen
  der Physik}\ }\textbf {\bibinfo {volume} {369}},\ \bibinfo {pages} {253}
  (\bibinfo {year} {1921})}\BibitemShut {NoStop}%
\bibitem [{\citenamefont {Hong}\ and\ \citenamefont
  {Virkar}(1995)}]{HongAndVirkar_JACS1995}%
  \BibitemOpen
  \bibfield  {author} {\bibinfo {author} {\bibfnamefont {S.~J.}\ \bibnamefont
  {Hong}}\ and\ \bibinfo {author} {\bibfnamefont {A.~V.}\ \bibnamefont
  {Virkar}},\ }\href@noop {} {\bibfield  {journal} {\bibinfo  {journal} {J Am
  Ceram Soc}\ }\textbf {\bibinfo {volume} {78}},\ \bibinfo {pages} {433}
  (\bibinfo {year} {1995})}\BibitemShut {NoStop}%
\bibitem [{\citenamefont {Grover}\ \emph {et~al.}(2008)\citenamefont {Grover},
  \citenamefont {Banerji}, \citenamefont {Sengupta},\ and\ \citenamefont
  {Tyagi}}]{GroverEtAl_JSSC2008}%
  \BibitemOpen
  \bibfield  {author} {\bibinfo {author} {\bibfnamefont {V.}~\bibnamefont
  {Grover}}, \bibinfo {author} {\bibfnamefont {A.}~\bibnamefont {Banerji}},
  \bibinfo {author} {\bibfnamefont {P.}~\bibnamefont {Sengupta}}, \ and\
  \bibinfo {author} {\bibfnamefont {A.}~\bibnamefont {Tyagi}},\ }\href@noop {}
  {\bibfield  {journal} {\bibinfo  {journal} {Journal of Solid State
  Chemistry}\ }\textbf {\bibinfo {volume} {181}},\ \bibinfo {pages} {1930 }
  (\bibinfo {year} {2008})}\BibitemShut {NoStop}%
\bibitem [{\citenamefont {Lee}\ \emph {et~al.}(2012)\citenamefont {Lee},
  \citenamefont {Meng}, \citenamefont {Kaneko},\ and\ \citenamefont
  {Tamaura}}]{LeeEtAl_JSEE2012}%
  \BibitemOpen
  \bibfield  {author} {\bibinfo {author} {\bibfnamefont {C.-i.}\ \bibnamefont
  {Lee}}, \bibinfo {author} {\bibfnamefont {Q.-L.}\ \bibnamefont {Meng}},
  \bibinfo {author} {\bibfnamefont {H.}~\bibnamefont {Kaneko}}, \ and\ \bibinfo
  {author} {\bibfnamefont {Y.}~\bibnamefont {Tamaura}},\ }\href@noop {}
  {\bibfield  {journal} {\bibinfo  {journal} {Journal of Solar Energy
  Engineering}\ }\textbf {\bibinfo {volume} {135}},\ \bibinfo {pages} {011002}
  (\bibinfo {year} {2012})}\BibitemShut {NoStop}%
\bibitem [{\citenamefont {Gerhardt-Anderson}\ and\ \citenamefont
  {Nowick}(1981)}]{Gerhardt-AndersonAndNowick_SSI1981}%
  \BibitemOpen
  \bibfield  {author} {\bibinfo {author} {\bibfnamefont {R.}~\bibnamefont
  {Gerhardt-Anderson}}\ and\ \bibinfo {author} {\bibfnamefont {A.~S.}\
  \bibnamefont {Nowick}},\ }\href@noop {} {\bibfield  {journal} {\bibinfo
  {journal} {Solid State Ionics}\ }\textbf {\bibinfo {volume} {5}},\ \bibinfo
  {pages} {547 } (\bibinfo {year} {1981})}\BibitemShut {NoStop}%
\bibitem [{\citenamefont {Huang}\ \emph {et~al.}(1998)\citenamefont {Huang},
  \citenamefont {Feng},\ and\ \citenamefont {Goodenough}}]{HuangEtAl_JACS1998}%
  \BibitemOpen
  \bibfield  {author} {\bibinfo {author} {\bibfnamefont {K.}~\bibnamefont
  {Huang}}, \bibinfo {author} {\bibfnamefont {M.}~\bibnamefont {Feng}}, \ and\
  \bibinfo {author} {\bibfnamefont {J.~B.}\ \bibnamefont {Goodenough}},\
  }\href@noop {} {\bibfield  {journal} {\bibinfo  {journal} {J Am Ceram Soc}\
  }\textbf {\bibinfo {volume} {81}},\ \bibinfo {pages} {357} (\bibinfo {year}
  {1998})}\BibitemShut {NoStop}%
\bibitem [{\citenamefont {Zhang}\ \emph {et~al.}(2002)\citenamefont {Zhang},
  \citenamefont {Hing}, \citenamefont {Huang},\ and\ \citenamefont
  {Kilner}}]{ZhangEtAl_SSI2002}%
  \BibitemOpen
  \bibfield  {author} {\bibinfo {author} {\bibfnamefont {T.}~\bibnamefont
  {Zhang}}, \bibinfo {author} {\bibfnamefont {P.}~\bibnamefont {Hing}},
  \bibinfo {author} {\bibfnamefont {H.}~\bibnamefont {Huang}}, \ and\ \bibinfo
  {author} {\bibfnamefont {J.}~\bibnamefont {Kilner}},\ }\href@noop {}
  {\bibfield  {journal} {\bibinfo  {journal} {Solid State Ionics}\ }\textbf
  {\bibinfo {volume} {148}},\ \bibinfo {pages} {567 } (\bibinfo {year}
  {2002})}\BibitemShut {NoStop}%
\bibitem [{\citenamefont {Huang}\ \emph {et~al.}(2011)\citenamefont {Huang},
  \citenamefont {Wei}, \citenamefont {Chen},\ and\ \citenamefont
  {Chen}}]{HuangEtAl_JECS2011}%
  \BibitemOpen
  \bibfield  {author} {\bibinfo {author} {\bibfnamefont {C.}~\bibnamefont
  {Huang}}, \bibinfo {author} {\bibfnamefont {W.}~\bibnamefont {Wei}}, \bibinfo
  {author} {\bibfnamefont {C.}~\bibnamefont {Chen}}, \ and\ \bibinfo {author}
  {\bibfnamefont {J.}~\bibnamefont {Chen}},\ }\href@noop {} {\bibfield
  {journal} {\bibinfo  {journal} {Journal of the European Ceramic Society}\
  }\textbf {\bibinfo {volume} {31}},\ \bibinfo {pages} {3159 } (\bibinfo {year}
  {2011})}\BibitemShut {NoStop}%
\bibitem [{\citenamefont {Buyukkilic}\ \emph {et~al.}(2012)\citenamefont
  {Buyukkilic}, \citenamefont {Shvareva},\ and\ \citenamefont
  {Navrotsky}}]{BuyukkilicEtAl_SSI2012}%
  \BibitemOpen
  \bibfield  {author} {\bibinfo {author} {\bibfnamefont {S.}~\bibnamefont
  {Buyukkilic}}, \bibinfo {author} {\bibfnamefont {T.}~\bibnamefont
  {Shvareva}}, \ and\ \bibinfo {author} {\bibfnamefont {A.}~\bibnamefont
  {Navrotsky}},\ }\href@noop {} {\bibfield  {journal} {\bibinfo  {journal}
  {Solid State Ionics}\ }\textbf {\bibinfo {volume} {227}},\ \bibinfo {pages}
  {17 } (\bibinfo {year} {2012})}\BibitemShut {NoStop}%
\bibitem [{\citenamefont {Hisashige}\ \emph {et~al.}(2006)\citenamefont
  {Hisashige}, \citenamefont {Yamamura},\ and\ \citenamefont
  {Tsuji}}]{HisashigeEtAl_JAC2006}%
  \BibitemOpen
  \bibfield  {author} {\bibinfo {author} {\bibfnamefont {T.}~\bibnamefont
  {Hisashige}}, \bibinfo {author} {\bibfnamefont {Y.}~\bibnamefont {Yamamura}},
  \ and\ \bibinfo {author} {\bibfnamefont {T.}~\bibnamefont {Tsuji}},\
  }\href@noop {} {\bibfield  {journal} {\bibinfo  {journal} {J Alloys Compd}\
  }\textbf {\bibinfo {volume} {408-412}},\ \bibinfo {pages} {1153 } (\bibinfo
  {year} {2006})}\BibitemShut {NoStop}%
\bibitem [{\citenamefont {Dikmen}\ \emph {et~al.}(1999)\citenamefont {Dikmen},
  \citenamefont {Shuk},\ and\ \citenamefont {Greenblatt}}]{DikmenEtAl_SSI1999}%
  \BibitemOpen
  \bibfield  {author} {\bibinfo {author} {\bibfnamefont {S.}~\bibnamefont
  {Dikmen}}, \bibinfo {author} {\bibfnamefont {P.}~\bibnamefont {Shuk}}, \ and\
  \bibinfo {author} {\bibfnamefont {M.}~\bibnamefont {Greenblatt}},\
  }\href@noop {} {\bibfield  {journal} {\bibinfo  {journal} {Solid State
  Ionics}\ }\textbf {\bibinfo {volume} {126}},\ \bibinfo {pages} {89 }
  (\bibinfo {year} {1999})}\BibitemShut {NoStop}%
\bibitem [{\citenamefont {Grieshammer}\ \emph {et~al.}(2014)\citenamefont
  {Grieshammer}, \citenamefont {Grope}, \citenamefont {Koettgen},\ and\
  \citenamefont {Martin}}]{GrieshammerEtAl_PCCP2014}%
  \BibitemOpen
  \bibfield  {author} {\bibinfo {author} {\bibfnamefont {S.}~\bibnamefont
  {Grieshammer}}, \bibinfo {author} {\bibfnamefont {B.~O.~H.}\ \bibnamefont
  {Grope}}, \bibinfo {author} {\bibfnamefont {J.}~\bibnamefont {Koettgen}}, \
  and\ \bibinfo {author} {\bibfnamefont {M.}~\bibnamefont {Martin}},\ }\href
  {\doibase 10.1039/C3CP54811B} {\bibfield  {journal} {\bibinfo  {journal}
  {Phys Chem Chem Phys}\ ,\ } (\bibinfo {year} {2014})}\BibitemShut {NoStop}%
\bibitem [{\citenamefont {Steele}(2000)}]{Steele_SSI2000}%
  \BibitemOpen
  \bibfield  {author} {\bibinfo {author} {\bibfnamefont {B.~C.~H.}\
  \bibnamefont {Steele}},\ }\href@noop {} {\bibfield  {journal} {\bibinfo
  {journal} {Solid State Ionics}\ }\textbf {\bibinfo {volume} {129}},\ \bibinfo
  {pages} {95 } (\bibinfo {year} {2000})}\BibitemShut {NoStop}%
\bibitem [{\citenamefont {Zhou}\ \emph {et~al.}(2002)\citenamefont {Zhou},
  \citenamefont {Huebner}, \citenamefont {Kosacki},\ and\ \citenamefont
  {Anderson}}]{ZhouEtAl_JACS2002}%
  \BibitemOpen
  \bibfield  {author} {\bibinfo {author} {\bibfnamefont {X.-D.}\ \bibnamefont
  {Zhou}}, \bibinfo {author} {\bibfnamefont {W.}~\bibnamefont {Huebner}},
  \bibinfo {author} {\bibfnamefont {I.}~\bibnamefont {Kosacki}}, \ and\
  \bibinfo {author} {\bibfnamefont {H.~U.}\ \bibnamefont {Anderson}},\
  }\href@noop {} {\bibfield  {journal} {\bibinfo  {journal} {J Am Ceram Soc}\
  }\textbf {\bibinfo {volume} {85}},\ \bibinfo {pages} {1757} (\bibinfo {year}
  {2002})}\BibitemShut {NoStop}%
\bibitem [{\citenamefont {Xia}\ and\ \citenamefont
  {Liu}(2002)}]{XiaAndLiu_SSI2002}%
  \BibitemOpen
  \bibfield  {author} {\bibinfo {author} {\bibfnamefont {C.}~\bibnamefont
  {Xia}}\ and\ \bibinfo {author} {\bibfnamefont {M.}~\bibnamefont {Liu}},\
  }\href@noop {} {\bibfield  {journal} {\bibinfo  {journal} {Solid State
  Ionics}\ }\textbf {\bibinfo {volume} {152-153}},\ \bibinfo {pages} {423 }
  (\bibinfo {year} {2002})}\BibitemShut {NoStop}%
\bibitem [{\citenamefont {Dholabhai}\ \emph {et~al.}(2012)\citenamefont
  {Dholabhai}, \citenamefont {Anwar}, \citenamefont {Adams}, \citenamefont
  {Crozier},\ and\ \citenamefont {Sharma}}]{DholabhaiEtAl_MSMSE2012}%
  \BibitemOpen
  \bibfield  {author} {\bibinfo {author} {\bibfnamefont {P.~P.}\ \bibnamefont
  {Dholabhai}}, \bibinfo {author} {\bibfnamefont {S.}~\bibnamefont {Anwar}},
  \bibinfo {author} {\bibfnamefont {J.~B.}\ \bibnamefont {Adams}}, \bibinfo
  {author} {\bibfnamefont {P.~A.}\ \bibnamefont {Crozier}}, \ and\ \bibinfo
  {author} {\bibfnamefont {R.}~\bibnamefont {Sharma}},\ }\href@noop {}
  {\bibfield  {journal} {\bibinfo  {journal} {Modelling and Simulation in
  Materials Science and Engineering}\ }\textbf {\bibinfo {volume} {20}},\
  \bibinfo {pages} {015004} (\bibinfo {year} {2012})}\BibitemShut {NoStop}%
\bibitem [{\citenamefont {Grope}\ \emph {et~al.}(2012)\citenamefont {Grope},
  \citenamefont {Zacherle}, \citenamefont {Nakayama},\ and\ \citenamefont
  {Martin}}]{GropeEtAl_SSI2012}%
  \BibitemOpen
  \bibfield  {author} {\bibinfo {author} {\bibfnamefont {B.}~\bibnamefont
  {Grope}}, \bibinfo {author} {\bibfnamefont {T.}~\bibnamefont {Zacherle}},
  \bibinfo {author} {\bibfnamefont {M.}~\bibnamefont {Nakayama}}, \ and\
  \bibinfo {author} {\bibfnamefont {M.}~\bibnamefont {Martin}},\ }\href@noop {}
  {\bibfield  {journal} {\bibinfo  {journal} {Solid State Ionics}\ }\textbf
  {\bibinfo {volume} {225}},\ \bibinfo {pages} {476 } (\bibinfo {year}
  {2012})}\BibitemShut {NoStop}%
\bibitem [{\citenamefont {Shemilt}\ and\ \citenamefont
  {Williams}(1999)}]{ShemiltAndWilliams_JMSL1999}%
  \BibitemOpen
  \bibfield  {author} {\bibinfo {author} {\bibfnamefont {J.}~\bibnamefont
  {Shemilt}}\ and\ \bibinfo {author} {\bibfnamefont {H.}~\bibnamefont
  {Williams}},\ }\href@noop {} {\bibfield  {journal} {\bibinfo  {journal}
  {Journal of Materials Science Letters}\ }\textbf {\bibinfo {volume} {18}},\
  \bibinfo {pages} {1735} (\bibinfo {year} {1999})}\BibitemShut {NoStop}%
\bibitem [{\citenamefont {Jung}\ \emph {et~al.}(2002)\citenamefont {Jung},
  \citenamefont {Huang},\ and\ \citenamefont {Chang}}]{JungEtAl_JSSE2002}%
  \BibitemOpen
  \bibfield  {author} {\bibinfo {author} {\bibfnamefont {G.-B.}\ \bibnamefont
  {Jung}}, \bibinfo {author} {\bibfnamefont {T.-J.}\ \bibnamefont {Huang}}, \
  and\ \bibinfo {author} {\bibfnamefont {C.-L.}\ \bibnamefont {Chang}},\
  }\href@noop {} {\bibfield  {journal} {\bibinfo  {journal} {J Solid State
  Electrochem}\ }\textbf {\bibinfo {volume} {6}},\ \bibinfo {pages} {225}
  (\bibinfo {year} {2002})}\BibitemShut {NoStop}%
\bibitem [{\citenamefont {Aneflous}\ \emph {et~al.}(2004)\citenamefont
  {Aneflous}, \citenamefont {Musso}, \citenamefont {Villain}, \citenamefont
  {Gavarri},\ and\ \citenamefont {Benyaich}}]{AneflousEtAl_JSSC2004}%
  \BibitemOpen
  \bibfield  {author} {\bibinfo {author} {\bibfnamefont {L.}~\bibnamefont
  {Aneflous}}, \bibinfo {author} {\bibfnamefont {J.~A.}\ \bibnamefont {Musso}},
  \bibinfo {author} {\bibfnamefont {S.}~\bibnamefont {Villain}}, \bibinfo
  {author} {\bibfnamefont {J.-R.}\ \bibnamefont {Gavarri}}, \ and\ \bibinfo
  {author} {\bibfnamefont {H.}~\bibnamefont {Benyaich}},\ }\href@noop {}
  {\bibfield  {journal} {\bibinfo  {journal} {Journal of Solid State
  Chemistry}\ }\textbf {\bibinfo {volume} {177}},\ \bibinfo {pages} {856 }
  (\bibinfo {year} {2004})}\BibitemShut {NoStop}%
\bibitem [{\citenamefont {Zhang}\ \emph {et~al.}(2004)\citenamefont {Zhang},
  \citenamefont {Ma}, \citenamefont {Kong}, \citenamefont {Chan},\ and\
  \citenamefont {Kilner}}]{ZhangEtAl_SSI2004}%
  \BibitemOpen
  \bibfield  {author} {\bibinfo {author} {\bibfnamefont {T.~S.}\ \bibnamefont
  {Zhang}}, \bibinfo {author} {\bibfnamefont {J.}~\bibnamefont {Ma}}, \bibinfo
  {author} {\bibfnamefont {L.~B.}\ \bibnamefont {Kong}}, \bibinfo {author}
  {\bibfnamefont {S.~H.}\ \bibnamefont {Chan}}, \ and\ \bibinfo {author}
  {\bibfnamefont {J.~A.}\ \bibnamefont {Kilner}},\ }\href@noop {} {\bibfield
  {journal} {\bibinfo  {journal} {Solid State Ionics}\ }\textbf {\bibinfo
  {volume} {170}},\ \bibinfo {pages} {209 } (\bibinfo {year}
  {2004})}\BibitemShut {NoStop}%
\bibitem [{\citenamefont {Stephens}\ and\ \citenamefont
  {Kilner}(2006)}]{StephensAndKilner_SSI2006}%
  \BibitemOpen
  \bibfield  {author} {\bibinfo {author} {\bibfnamefont {I.~E.}\ \bibnamefont
  {Stephens}}\ and\ \bibinfo {author} {\bibfnamefont {J.~A.}\ \bibnamefont
  {Kilner}},\ }\href@noop {} {\bibfield  {journal} {\bibinfo  {journal} {Solid
  State Ionics}\ }\textbf {\bibinfo {volume} {177}},\ \bibinfo {pages} {669 }
  (\bibinfo {year} {2006})}\BibitemShut {NoStop}%
\bibitem [{\citenamefont {Lang}\ \emph {et~al.}(1999)\citenamefont {Lang},
  \citenamefont {Kiinstler}, \citenamefont {Mangier},\ and\ \citenamefont
  {Tomandl}}]{LangEtAl_JES1999}%
  \BibitemOpen
  \bibfield  {author} {\bibinfo {author} {\bibfnamefont {H.}~\bibnamefont
  {Lang}}, \bibinfo {author} {\bibfnamefont {K.}~\bibnamefont {Kiinstler}},
  \bibinfo {author} {\bibfnamefont {M.}~\bibnamefont {Mangier}}, \ and\
  \bibinfo {author} {\bibfnamefont {G.}~\bibnamefont {Tomandl}},\ }in\
  \href@noop {} {\emph {\bibinfo {booktitle} {Solid Oxide Fuel Cells (SOFC VI):
  Proceedings of the Sixth International Symposium}}},\ Vol.~\bibinfo {volume}
  {6}\ (\bibinfo {year} {1999})\ p.\ \bibinfo {pages} {233}\BibitemShut
  {NoStop}%
\bibitem [{\citenamefont {Li}\ \emph {et~al.}(1991)\citenamefont {Li},
  \citenamefont {Chen}, \citenamefont {Penner-Hahn},\ and\ \citenamefont
  {Tien}}]{LiEtAl_JACS1991}%
  \BibitemOpen
  \bibfield  {author} {\bibinfo {author} {\bibfnamefont {P.}~\bibnamefont
  {Li}}, \bibinfo {author} {\bibfnamefont {I.-W.}\ \bibnamefont {Chen}},
  \bibinfo {author} {\bibfnamefont {J.~E.}\ \bibnamefont {Penner-Hahn}}, \ and\
  \bibinfo {author} {\bibfnamefont {T.-Y.}\ \bibnamefont {Tien}},\ }\href@noop
  {} {\bibfield  {journal} {\bibinfo  {journal} {J Am Ceram Soc}\ }\textbf
  {\bibinfo {volume} {74}},\ \bibinfo {pages} {958} (\bibinfo {year}
  {1991})}\BibitemShut {NoStop}%
\bibitem [{\citenamefont {Nakayama}\ and\ \citenamefont
  {Martin}(2009)}]{NakayamaAndMartin_PCCP2009}%
  \BibitemOpen
  \bibfield  {author} {\bibinfo {author} {\bibfnamefont {M.}~\bibnamefont
  {Nakayama}}\ and\ \bibinfo {author} {\bibfnamefont {M.}~\bibnamefont
  {Martin}},\ }\href@noop {} {\bibfield  {journal} {\bibinfo  {journal} {Phys
  Chem Chem Phys}\ }\textbf {\bibinfo {volume} {11}},\ \bibinfo {pages} {3241}
  (\bibinfo {year} {2009})}\BibitemShut {NoStop}%
\bibitem [{\citenamefont {Hull}\ \emph {et~al.}(2009)\citenamefont {Hull},
  \citenamefont {Norberg}, \citenamefont {Ahmed}, \citenamefont {Eriksson},
  \citenamefont {Marrocchelli},\ and\ \citenamefont
  {Madden}}]{HullEtAl_JSSC2009}%
  \BibitemOpen
  \bibfield  {author} {\bibinfo {author} {\bibfnamefont {S.}~\bibnamefont
  {Hull}}, \bibinfo {author} {\bibfnamefont {S.~T.}\ \bibnamefont {Norberg}},
  \bibinfo {author} {\bibfnamefont {I.}~\bibnamefont {Ahmed}}, \bibinfo
  {author} {\bibfnamefont {S.~G.}\ \bibnamefont {Eriksson}}, \bibinfo {author}
  {\bibfnamefont {D.}~\bibnamefont {Marrocchelli}}, \ and\ \bibinfo {author}
  {\bibfnamefont {P.~A.}\ \bibnamefont {Madden}},\ }\href@noop {} {\bibfield
  {journal} {\bibinfo  {journal} {J Solid State Chem}\ }\textbf {\bibinfo
  {volume} {182}},\ \bibinfo {pages} {2815 } (\bibinfo {year}
  {2009})}\BibitemShut {NoStop}%
\bibitem [{\citenamefont {Gopal}\ and\ \citenamefont {van~de
  Walle}(2012)}]{GopalAndvandeWalle_PRB2012}%
  \BibitemOpen
  \bibfield  {author} {\bibinfo {author} {\bibfnamefont {C.~B.}\ \bibnamefont
  {Gopal}}\ and\ \bibinfo {author} {\bibfnamefont {A.}~\bibnamefont {van~de
  Walle}},\ }\href@noop {} {\bibfield  {journal} {\bibinfo  {journal} {Physical
  Review B}\ }\textbf {\bibinfo {volume} {86}},\ \bibinfo {pages} {134117}
  (\bibinfo {year} {2012})}\BibitemShut {NoStop}%
\bibitem [{\citenamefont {Dikmen}(2010)}]{Dikmen_JAC2010}%
  \BibitemOpen
  \bibfield  {author} {\bibinfo {author} {\bibfnamefont {S.}~\bibnamefont
  {Dikmen}},\ }\href@noop {} {\bibfield  {journal} {\bibinfo  {journal} {J
  Alloys Compd}\ }\textbf {\bibinfo {volume} {491}},\ \bibinfo {pages} {106 }
  (\bibinfo {year} {2010})},\ \bibinfo {note} {ceO2:77}\BibitemShut {NoStop}%
\bibitem [{\citenamefont {Wang}\ \emph {et~al.}(2004)\citenamefont {Wang},
  \citenamefont {Chen},\ and\ \citenamefont {Cheng}}]{WangEtAl_EC2004}%
  \BibitemOpen
  \bibfield  {author} {\bibinfo {author} {\bibfnamefont {F.-Y.}\ \bibnamefont
  {Wang}}, \bibinfo {author} {\bibfnamefont {S.}~\bibnamefont {Chen}}, \ and\
  \bibinfo {author} {\bibfnamefont {S.}~\bibnamefont {Cheng}},\ }\href@noop {}
  {\bibfield  {journal} {\bibinfo  {journal} {Electrochemistry Communications}\
  }\textbf {\bibinfo {volume} {6}},\ \bibinfo {pages} {743 } (\bibinfo {year}
  {2004})}\BibitemShut {NoStop}%
\bibitem [{\citenamefont {Yoshida}\ \emph {et~al.}(2001)\citenamefont
  {Yoshida}, \citenamefont {Deguchi}, \citenamefont {Miura}, \citenamefont
  {Horiuchi},\ and\ \citenamefont {Inagaki}}]{YoshidaEtAl_SSI2001}%
  \BibitemOpen
  \bibfield  {author} {\bibinfo {author} {\bibfnamefont {H.}~\bibnamefont
  {Yoshida}}, \bibinfo {author} {\bibfnamefont {H.}~\bibnamefont {Deguchi}},
  \bibinfo {author} {\bibfnamefont {K.}~\bibnamefont {Miura}}, \bibinfo
  {author} {\bibfnamefont {M.}~\bibnamefont {Horiuchi}}, \ and\ \bibinfo
  {author} {\bibfnamefont {T.}~\bibnamefont {Inagaki}},\ }\href@noop {}
  {\bibfield  {journal} {\bibinfo  {journal} {Solid State Ionics}\ }\textbf
  {\bibinfo {volume} {140}},\ \bibinfo {pages} {191 } (\bibinfo {year}
  {2001})}\BibitemShut {NoStop}%
\bibitem [{\citenamefont {Yoshida}\ \emph {et~al.}(2003)\citenamefont
  {Yoshida}, \citenamefont {Inagaki}, \citenamefont {Miura}, \citenamefont
  {Inaba},\ and\ \citenamefont {Ogumi}}]{YoshidaEtAl_SSI2003}%
  \BibitemOpen
  \bibfield  {author} {\bibinfo {author} {\bibfnamefont {H.}~\bibnamefont
  {Yoshida}}, \bibinfo {author} {\bibfnamefont {T.}~\bibnamefont {Inagaki}},
  \bibinfo {author} {\bibfnamefont {K.}~\bibnamefont {Miura}}, \bibinfo
  {author} {\bibfnamefont {M.}~\bibnamefont {Inaba}}, \ and\ \bibinfo {author}
  {\bibfnamefont {Z.}~\bibnamefont {Ogumi}},\ }\href@noop {} {\bibfield
  {journal} {\bibinfo  {journal} {Solid State Ionics}\ }\textbf {\bibinfo
  {volume} {160}},\ \bibinfo {pages} {109 } (\bibinfo {year}
  {2003})}\BibitemShut {NoStop}%
\bibitem [{\citenamefont {Norberg}\ \emph {et~al.}(2012)\citenamefont
  {Norberg}, \citenamefont {Hull}, \citenamefont {Eriksson}, \citenamefont
  {Ahmed}, \citenamefont {Kinyanjui},\ and\ \citenamefont
  {Biendicho}}]{NorbergEtAl_CM2013}%
  \BibitemOpen
  \bibfield  {author} {\bibinfo {author} {\bibfnamefont {S.~T.}\ \bibnamefont
  {Norberg}}, \bibinfo {author} {\bibfnamefont {S.}~\bibnamefont {Hull}},
  \bibinfo {author} {\bibfnamefont {S.~G.}\ \bibnamefont {Eriksson}}, \bibinfo
  {author} {\bibfnamefont {I.}~\bibnamefont {Ahmed}}, \bibinfo {author}
  {\bibfnamefont {F.}~\bibnamefont {Kinyanjui}}, \ and\ \bibinfo {author}
  {\bibfnamefont {J.~J.}\ \bibnamefont {Biendicho}},\ }\href@noop {} {\bibfield
   {journal} {\bibinfo  {journal} {Chem Mater}\ }\textbf {\bibinfo {volume}
  {24}},\ \bibinfo {pages} {4294} (\bibinfo {year} {2012})}\BibitemShut
  {NoStop}%
\bibitem [{\citenamefont {Kushima}\ and\ \citenamefont
  {Yildiz}(2009)}]{KushimaAndYildizECST2009}%
  \BibitemOpen
  \bibfield  {author} {\bibinfo {author} {\bibfnamefont {A.}~\bibnamefont
  {Kushima}}\ and\ \bibinfo {author} {\bibfnamefont {B.}~\bibnamefont
  {Yildiz}},\ }\href@noop {} {\bibfield  {journal} {\bibinfo  {journal} {ECS
  Transactions}\ }\textbf {\bibinfo {volume} {25}},\ \bibinfo {pages} {1599}
  (\bibinfo {year} {2009})}\BibitemShut {NoStop}%
\bibitem [{\citenamefont {Kushima}\ and\ \citenamefont
  {Yildiz}(2010)}]{KushimaAndYildiz_JMC2010}%
  \BibitemOpen
  \bibfield  {author} {\bibinfo {author} {\bibfnamefont {A.}~\bibnamefont
  {Kushima}}\ and\ \bibinfo {author} {\bibfnamefont {B.}~\bibnamefont
  {Yildiz}},\ }\href@noop {} {\bibfield  {journal} {\bibinfo  {journal} {J
  Mater Chem}\ }\textbf {\bibinfo {volume} {20}},\ \bibinfo {pages} {4809}
  (\bibinfo {year} {2010})}\BibitemShut {NoStop}%
\bibitem [{\citenamefont {Norberg}\ \emph {et~al.}(2009)\citenamefont
  {Norberg}, \citenamefont {Ahmed}, \citenamefont {Hull}, \citenamefont
  {Marrocchelli},\ and\ \citenamefont {Madden}}]{NorbergEtAl_JPCM2009}%
  \BibitemOpen
  \bibfield  {author} {\bibinfo {author} {\bibfnamefont {S.~T.}\ \bibnamefont
  {Norberg}}, \bibinfo {author} {\bibfnamefont {I.}~\bibnamefont {Ahmed}},
  \bibinfo {author} {\bibfnamefont {S.}~\bibnamefont {Hull}}, \bibinfo {author}
  {\bibfnamefont {D.}~\bibnamefont {Marrocchelli}}, \ and\ \bibinfo {author}
  {\bibfnamefont {P.~A.}\ \bibnamefont {Madden}},\ }\href@noop {} {\bibfield
  {journal} {\bibinfo  {journal} {J Phys : Condens Matter}\ }\textbf {\bibinfo
  {volume} {21}},\ \bibinfo {pages} {215401} (\bibinfo {year} {2009})},\
  \bibinfo {note} {ceO2:10}\BibitemShut {NoStop}%
\bibitem [{\citenamefont {Madden}\ \emph {et~al.}(2006)\citenamefont {Madden},
  \citenamefont {Heaton}, \citenamefont {Aguado},\ and\ \citenamefont
  {Jahn}}]{MaddenEtAl_JMST2006}%
  \BibitemOpen
  \bibfield  {author} {\bibinfo {author} {\bibfnamefont {P.~A.}\ \bibnamefont
  {Madden}}, \bibinfo {author} {\bibfnamefont {R.}~\bibnamefont {Heaton}},
  \bibinfo {author} {\bibfnamefont {A.}~\bibnamefont {Aguado}}, \ and\ \bibinfo
  {author} {\bibfnamefont {S.}~\bibnamefont {Jahn}},\ }\href@noop {} {\bibfield
   {journal} {\bibinfo  {journal} {Journal of Molecular Structure: THEOCHEM}\
  }\textbf {\bibinfo {volume} {771}},\ \bibinfo {pages} {9} (\bibinfo {year}
  {2006})}\BibitemShut {NoStop}%
\bibitem [{\citenamefont {Castiglione}\ \emph {et~al.}(1999)\citenamefont
  {Castiglione}, \citenamefont {Wilson},\ and\ \citenamefont
  {Madden}}]{CastiglioneEtSAl_JPCM1999}%
  \BibitemOpen
  \bibfield  {author} {\bibinfo {author} {\bibfnamefont {M.~J.}\ \bibnamefont
  {Castiglione}}, \bibinfo {author} {\bibfnamefont {M.}~\bibnamefont {Wilson}},
  \ and\ \bibinfo {author} {\bibfnamefont {P.~A.}\ \bibnamefont {Madden}},\
  }\href@noop {} {\bibfield  {journal} {\bibinfo  {journal} {J Phys : Condens
  Matter}\ }\textbf {\bibinfo {volume} {11}},\ \bibinfo {pages} {9009}
  (\bibinfo {year} {1999})},\ \bibinfo {note} {mD:66}\BibitemShut {NoStop}%
\bibitem [{\citenamefont {Tang}\ and\ \citenamefont
  {Toennies}(1984)}]{TangAndToennies_JCP1984}%
  \BibitemOpen
  \bibfield  {author} {\bibinfo {author} {\bibfnamefont {K.~T.}\ \bibnamefont
  {Tang}}\ and\ \bibinfo {author} {\bibfnamefont {J.~P.}\ \bibnamefont
  {Toennies}},\ }\href@noop {} {\bibfield  {journal} {\bibinfo  {journal} {The
  Journal of Chemical Physics}\ }\textbf {\bibinfo {volume} {80}},\ \bibinfo
  {pages} {3726} (\bibinfo {year} {1984})}\BibitemShut {NoStop}%
\bibitem [{\citenamefont {Tang}\ and\ \citenamefont
  {Toennies}(2003)}]{TangAndToennies_JCP2003}%
  \BibitemOpen
  \bibfield  {author} {\bibinfo {author} {\bibfnamefont {K.~T.}\ \bibnamefont
  {Tang}}\ and\ \bibinfo {author} {\bibfnamefont {J.~P.}\ \bibnamefont
  {Toennies}},\ }\href@noop {} {\bibfield  {journal} {\bibinfo  {journal} {The
  Journal of Chemical Physics}\ }\textbf {\bibinfo {volume} {118}},\ \bibinfo
  {pages} {4976} (\bibinfo {year} {2003})}\BibitemShut {NoStop}%
\bibitem [{\citenamefont {Marzari}\ and\ \citenamefont
  {Vanderbilt}(1997)}]{MarzariAndVanderbilt_PRB1997}%
  \BibitemOpen
  \bibfield  {author} {\bibinfo {author} {\bibfnamefont {N.}~\bibnamefont
  {Marzari}}\ and\ \bibinfo {author} {\bibfnamefont {D.}~\bibnamefont
  {Vanderbilt}},\ }\href@noop {} {\bibfield  {journal} {\bibinfo  {journal}
  {Physical Review B}\ }\textbf {\bibinfo {volume} {56}},\ \bibinfo {pages}
  {12847} (\bibinfo {year} {1997})},\ \bibinfo {note} {wannier:1}\BibitemShut
  {NoStop}%
\bibitem [{\citenamefont {Heyd}\ \emph {et~al.}(2003)\citenamefont {Heyd},
  \citenamefont {Scuseria},\ and\ \citenamefont
  {Ernzerhof}}]{HeydEtAl_JCP2003}%
  \BibitemOpen
  \bibfield  {author} {\bibinfo {author} {\bibfnamefont {J.}~\bibnamefont
  {Heyd}}, \bibinfo {author} {\bibfnamefont {G.~E.}\ \bibnamefont {Scuseria}},
  \ and\ \bibinfo {author} {\bibfnamefont {M.}~\bibnamefont {Ernzerhof}},\
  }\href@noop {} {\bibfield  {journal} {\bibinfo  {journal} {J Chem Phys}\
  }\textbf {\bibinfo {volume} {118}},\ \bibinfo {pages} {8207} (\bibinfo {year}
  {2003})}\BibitemShut {NoStop}%
\bibitem [{\citenamefont {Krukau}\ \emph {et~al.}(2006)\citenamefont {Krukau},
  \citenamefont {Vydrov}, \citenamefont {Izmaylov},\ and\ \citenamefont
  {Scuseria}}]{KrukauEtAl_JCP2006}%
  \BibitemOpen
  \bibfield  {author} {\bibinfo {author} {\bibfnamefont {A.~V.}\ \bibnamefont
  {Krukau}}, \bibinfo {author} {\bibfnamefont {O.~A.}\ \bibnamefont {Vydrov}},
  \bibinfo {author} {\bibfnamefont {A.~F.}\ \bibnamefont {Izmaylov}}, \ and\
  \bibinfo {author} {\bibfnamefont {G.~E.}\ \bibnamefont {Scuseria}},\
  }\href@noop {} {\bibfield  {journal} {\bibinfo  {journal} {J Chem Phys}\
  }\textbf {\bibinfo {volume} {125}},\ \bibinfo {pages} {224106} (\bibinfo
  {year} {2006})}\BibitemShut {NoStop}%
\bibitem [{\citenamefont {Paier}\ \emph {et~al.}(2006)\citenamefont {Paier},
  \citenamefont {Marsman}, \citenamefont {Hummer}, \citenamefont {Kresse},
  \citenamefont {Gerber},\ and\ \citenamefont {Angyan}}]{PaierEtAl_JCP2006}%
  \BibitemOpen
  \bibfield  {author} {\bibinfo {author} {\bibfnamefont {J.}~\bibnamefont
  {Paier}}, \bibinfo {author} {\bibfnamefont {M.}~\bibnamefont {Marsman}},
  \bibinfo {author} {\bibfnamefont {K.}~\bibnamefont {Hummer}}, \bibinfo
  {author} {\bibfnamefont {G.}~\bibnamefont {Kresse}}, \bibinfo {author}
  {\bibfnamefont {I.~C.}\ \bibnamefont {Gerber}}, \ and\ \bibinfo {author}
  {\bibfnamefont {J.~G.}\ \bibnamefont {Angyan}},\ }\href@noop {} {\bibfield
  {journal} {\bibinfo  {journal} {The Journal of Chemical Physics}\ }\textbf
  {\bibinfo {volume} {124}},\ \bibinfo {pages} {154709} (\bibinfo {year}
  {2006})}\BibitemShut {NoStop}%
\bibitem [{\citenamefont {Koch}\ and\ \citenamefont
  {Holthausen}(2001)}]{KochEtAndHolthausen_Book2001}%
  \BibitemOpen
  \bibfield  {author} {\bibinfo {author} {\bibfnamefont {W.}~\bibnamefont
  {Koch}}\ and\ \bibinfo {author} {\bibfnamefont {M.~C.}\ \bibnamefont
  {Holthausen}},\ }\href@noop {} {\emph {\bibinfo {title} {{A Chemist's Guide
  to Density Functional Theory}}}}\ (\bibinfo  {publisher} {Wiley-VCH Verlag
  GmbH, Weinheim, Germany},\ \bibinfo {year} {2001})\BibitemShut {NoStop}%
\bibitem [{\citenamefont {Heaton}\ \emph
  {et~al.}(2006{\natexlab{b}})\citenamefont {Heaton}, \citenamefont {Madden},
  \citenamefont {Clark},\ and\ \citenamefont {Jahn}}]{HeatonEtAl_JCP2006}%
  \BibitemOpen
  \bibfield  {author} {\bibinfo {author} {\bibfnamefont {R.~J.}\ \bibnamefont
  {Heaton}}, \bibinfo {author} {\bibfnamefont {P.~A.}\ \bibnamefont {Madden}},
  \bibinfo {author} {\bibfnamefont {S.~J.}\ \bibnamefont {Clark}}, \ and\
  \bibinfo {author} {\bibfnamefont {S.}~\bibnamefont {Jahn}},\ }\href@noop {}
  {\bibfield  {journal} {\bibinfo  {journal} {The Journal of Chemical Physics}\
  }\textbf {\bibinfo {volume} {125}},\ \bibinfo {pages} {144104} (\bibinfo
  {year} {2006}{\natexlab{b}})}\BibitemShut {NoStop}%
\bibitem [{\citenamefont {Slater}\ and\ \citenamefont
  {Kirkwood}(1931)}]{SlaterAndKirkwood_PR1931}%
  \BibitemOpen
  \bibfield  {author} {\bibinfo {author} {\bibfnamefont {J.~C.}\ \bibnamefont
  {Slater}}\ and\ \bibinfo {author} {\bibfnamefont {J.~G.}\ \bibnamefont
  {Kirkwood}},\ }\href@noop {} {\bibfield  {journal} {\bibinfo  {journal} {Phys
  Rev}\ }\textbf {\bibinfo {volume} {37}},\ \bibinfo {pages} {682} (\bibinfo
  {year} {1931})}\BibitemShut {NoStop}%
\bibitem [{\citenamefont {Haigis}\ \emph {et~al.}(2013)\citenamefont {Haigis},
  \citenamefont {Salanne}, \citenamefont {Simon}, \citenamefont {Wilke},\ and\
  \citenamefont {Jahn}}]{HaigisEtAl_CG2013}%
  \BibitemOpen
  \bibfield  {author} {\bibinfo {author} {\bibfnamefont {V.}~\bibnamefont
  {Haigis}}, \bibinfo {author} {\bibfnamefont {M.}~\bibnamefont {Salanne}},
  \bibinfo {author} {\bibfnamefont {S.}~\bibnamefont {Simon}}, \bibinfo
  {author} {\bibfnamefont {M.}~\bibnamefont {Wilke}}, \ and\ \bibinfo {author}
  {\bibfnamefont {S.}~\bibnamefont {Jahn}},\ }\href@noop {} {\bibfield
  {journal} {\bibinfo  {journal} {Chemical Geology}\ }\textbf {\bibinfo
  {volume} {346}},\ \bibinfo {pages} {14 } (\bibinfo {year}
  {2013})}\BibitemShut {NoStop}%
\bibitem [{\citenamefont {Castiglione}\ and\ \citenamefont
  {Madden}(2001)}]{CastiglioneAndMadden_JPCM2001}%
  \BibitemOpen
  \bibfield  {author} {\bibinfo {author} {\bibfnamefont {M.~J.}\ \bibnamefont
  {Castiglione}}\ and\ \bibinfo {author} {\bibfnamefont {P.~A.}\ \bibnamefont
  {Madden}},\ }\href@noop {} {\bibfield  {journal} {\bibinfo  {journal} {J Phys
  : Condens Matter}\ }\textbf {\bibinfo {volume} {13}},\ \bibinfo {pages}
  {9963} (\bibinfo {year} {2001})}\BibitemShut {NoStop}%
\bibitem [{\citenamefont {Abrahams}\ \emph {et~al.}(2010)\citenamefont
  {Abrahams}, \citenamefont {Liu}, \citenamefont {Hull}, \citenamefont
  {Norberg}, \citenamefont {Krok}, \citenamefont {Kozanecka-Szmigiel},
  \citenamefont {Islam},\ and\ \citenamefont {Stokes}}]{AbrahamsEtAL_CM2010}%
  \BibitemOpen
  \bibfield  {author} {\bibinfo {author} {\bibfnamefont {I.}~\bibnamefont
  {Abrahams}}, \bibinfo {author} {\bibfnamefont {X.}~\bibnamefont {Liu}},
  \bibinfo {author} {\bibfnamefont {S.}~\bibnamefont {Hull}}, \bibinfo {author}
  {\bibfnamefont {S.~T.}\ \bibnamefont {Norberg}}, \bibinfo {author}
  {\bibfnamefont {F.}~\bibnamefont {Krok}}, \bibinfo {author} {\bibfnamefont
  {A.}~\bibnamefont {Kozanecka-Szmigiel}}, \bibinfo {author} {\bibfnamefont
  {M.~S.}\ \bibnamefont {Islam}}, \ and\ \bibinfo {author} {\bibfnamefont
  {S.~J.}\ \bibnamefont {Stokes}},\ }\href@noop {} {\bibfield  {journal}
  {\bibinfo  {journal} {Chem Mater}\ }\textbf {\bibinfo {volume} {22}},\
  \bibinfo {pages} {4435} (\bibinfo {year} {2010})}\BibitemShut {NoStop}%
\bibitem [{\citenamefont {Shannon}(1976)}]{Shannon_ACA1976}%
  \BibitemOpen
  \bibfield  {author} {\bibinfo {author} {\bibfnamefont {R.}~\bibnamefont
  {Shannon}},\ }\href@noop {} {\bibfield  {journal} {\bibinfo  {journal} {Acta
  Crystallographica Section A}\ }\textbf {\bibinfo {volume} {32}},\ \bibinfo
  {pages} {751} (\bibinfo {year} {1976})}\BibitemShut {NoStop}%
\end{thebibliography}

\end{document}